\documentclass[12pt]{article}
\usepackage{epsfig,amssymb}
\usepackage{latexsym}

\newcommand{\bq}{\begin{equation}}
\newcommand{\eq}{\end{equation}}
\newcommand{\bqa}{\begin{eqnarray}}
\newcommand{\eqa}{\end{eqnarray}}
\newcommand{\ben}{\begin{enumerate}}
\newcommand{\een}{\end{enumerate}}
\newcommand{\bc}{\begin{center}}
\newcommand{\ec}{\end{center}}
\newcommand{\bqb}{\begin{eqnarray*}}
\newcommand{\eqb}{\end{eqnarray*}}

\def\pr#1#2#3{ Phys. Rev. ${\bf{#1}}$:#2 (#3)}
\def\prl#1#2#3{ Phys. Rev. Lett. ${\bf{#1}}$:#2 (#3)}
\def\pl#1#2#3{ Phys. Lett. ${\bf{#1}}$:#2 (#3)}
\def\prep#1#2#3{ Phys. Rep. ${\bf{#1}}$:#2 (#3)}
\def\rmp#1#2#3{ Rev. Mod. Phys. ${\bf{#1}}$:#2(#3)}

\def\np#1#2#3{ Nucl. Phys. ${\bf{#1}}$:#2 (#3)}

\def\jhep#1#2#3{ JHEP ${\bf{#1}}$:#2 (#3)}
\def\epj#1#2#3{ Eur. Phys. J. ${\bf{#1}}$:#2 (#3)}

\def\astrojs#1#2#3{Astrophys. J. Suppl. ${\bf{#1}}$:#2 (#3)}

\def\polon#1#2#3{Acta Phys. Polon. ${\bf{#1}}$:#2 (#3) }

\def\ie{{\it i.e. }}
\def\eg{{\it e.g. }}

\def\etal{{\it et.al.}}

\global\nulldelimiterspace = 0pt

\def\d{\mathrm d}

\def\R{ {\cal R }}

\def\sw{s_W}
\def\cw{c_W}
\def\swd{s^2_W}
\def\cwd{c^2_W}

\def\mwd{m_W^2}

\def\mi{m_i}
\def\mid{m_i^2}
\def\mj{m_j}
\def\mjd{m_j^2}

\def\tchi{\tilde \chi}

\def\tg{\tilde g}
\def\s{\hat s}
\def\t{\hat t}
\def\u{\hat u}

\begin{document}
\pagenumbering{arabic}
\thispagestyle{empty}
\def\thefootnote{\fnsymbol{footnote}}
\setcounter{footnote}{1}

\begin{flushright}
November 29th,  2004\\
PM/04-38\\
hep-ph/0411366\\
corrected version, March 17th, 2005\\

 \end{flushright}
\vspace{2cm}
\begin{center}
{\Large\bf Single Neutralino production at CERN
LHC\footnote{Work supported by the European Union under the RTN contracts
HPRN-CT-2000-00149.}}.
\vspace{1.cm}  \\
{\large G.J. Gounaris$^a$, J. Layssac$^b$, P.I. Porfyriadis $^a$
and F.M. Renard$^b$}\\
\vspace{0.2cm}
$^a$Department of Theoretical Physics, Aristotle University of Thessaloniki,
\\ Gr-54124, Thessaloniki, Greece.\\
\vspace{0.2cm}
$^b$Physique Math\'{e}matique et Th\'{e}orique, UMR 5825\\
Universit\'{e} Montpellier II,  F-34095 Montpellier Cedex 5.
\vspace*{1.cm}

{\bf Abstract}
\end{center}
\vspace*{-0.4cm}
The common belief that the lightest supersymmetric particle (LSP)
might be a neutralino, providing also the main Dark Matter (DM)
component, calls for  maximal detail in the study of the neutralino
properties. Motivated by this, we  consider the direct production
of a single neutralino $\tchi^0_i$ at a high/energy hadron collider,
focusing on  the $\tchi^0_1$ and $\tchi^0_2$ cases.
At Born  level, the relevant subprocesses are
$q\bar q\to \tchi^0_i \tilde g$,
$g q\to \tchi^0_i \tilde q_{L,R}$ and
$q\bar q'\to \tchi^0_i\tchi^\pm_j$; while at 1-loop,
apart from radiative corrections to these processes,
we consider also $gg\to \tchi^0_i\tilde{g}$, for which a
numerical code named PLATONgluino is released. The relative importance
of these channels turns out to be extremely model dependent.
Combining these results  with an analogous study of the direct
$\tchi^0_i\tchi^0_j$ pair production, should help in  testing
the SUSY models and the Dark Matter assignment.

\vspace{0.5cm}

\def\thefootnote{\arabic{footnote}}
\setcounter{footnote}{0}
\clearpage

\section{Introduction}

The lightest neutralino state, $\tchi^0_1$, is often assumed
to be the Lightest Supersymmetric Particle (LSP) \cite{LSP}.
As such it is also a candidate for the origin of Dark Matter
(DM)  \cite{DMLSP}. This assumption has of course to be verified though, by analyzing the
results or  constraints reached by experiments trying to detect  Dark Matter
through  direct or indirect methods  \cite{DMobs, DMann}.

However, even in the minimal MSSM version of the   SUSY models,
the large parameter space induces   great   uncertainties in the neutralino
properties. So to check the consistency of the DM idea, it is  essential
to establish the neutralino  properties through    production
 at high energy hadron and  lepton  colliders.
The first such possibility of neutralino production, will probably be  through cascades at
the CERN LHC \cite{SUSYsearches1, SUSYsearches2}. But precious additional independent information
from LHC could also be obtained by studying  the smaller signals of  the
direct   ($\tchi^0_i \tchi^0_j$) pair production, as well as  the  production
in association with other   sparticles   in processes as $(\tchi^0_i \tilde g)$,
($\tchi^0_i \tilde q_{L,R}$) or ($\tchi^0_i \tchi^{\pm}_j$). When
the LC collider, will finally be built, a wealth of additional information will become
accessible \cite{gamma-gamma}.

Studies of the pure QCD effects to these channels   at LO
and  NLO have already appeared \cite{Spiratalk, Beenakker}.
 A recent summary can be found in
 \cite{Plehn}, where the results of a NLO QCD computation
are presented for various processes including neutralino production in association with
 a gluino, squark, slepton, chargino, or another neutralino.
The overall conclusion  of these computations is that at the LHC range,  the pure
QCD soft and collinear
corrections always {\it increase} the LO cross section   by an amount which,
depending on the subprocess c.m. energy and  the masses of the particles involved, lies
in the range of 10\% to 40\%.

Also important at  LHC though, turn out to be the
leading and subleading 1-loop logarithmic (LL) electroweak (EW) corrections.
Particularly for processes characterized by
 non-vanishing Born contributions, such effects show  a largely
universal structure with the leading $\ln^2(\hat s)$-terms solely
determined by the couplings of  the known gauge bosons $(W^\pm,Z,\gamma)$ to the
external particles of the  process; which in turn are  fixed completely by  their quantum numbers.
The situation is different for the subleading single-$\ln (\hat s)$ terms though, which
depend  on  the  couplings and masses of all
virtual particles, gauge or non-gauge,  shaping up the underlying dynamics
 \cite{BRV, Denner, qqLHC, Beenakker}.
Thus, depending on whether SUSY is  "near by"  with all
MSSM  sparticles below the TeV range,
or some of the sparticles are very heavy,
or even that the pure simple SM model
stays correct till very high scales, will only affect the subleading single $\ln(\hat s)$-terms
 \cite{BRV, qqLHC, gam-gam-VV}.

The most striking characteristic when comparing these EW corrections
to the aforementioned pure QCD ones, is that
 they are of roughly similar magnitude, but have  {\it opposite sign}
 \cite{BRV, Denner, qqLHC, gam-gam-VV}.

Particularly for the subprocess $q\bar q\to\tchi^0_i \tchi^0_j$ contributing
to the  neutralino-pair production, these effects have been studied in
\cite{pp-2chi0}, where the calculation of  the pure  1-loop
  process $gg\to\tchi^0_i \tchi^0_j$ was also included. If the masses
  of the squarks of the 1rst and 2nd family turn out to be very heavy,
  it might happen  at LHC, that the $gg\to\tchi^0_i \tchi^0_j$ contribution
  is comparable to  that of the LO process
  $q\bar q\to\tchi^0_i \tchi^0_j$, particularly at low invariant masses where
  the gluon flux is very large.\\

Additional information on neutralinos in a hadron collider could be obtained
from the single neutralino production triggered by the  subprocesses:
\bq
~~q\bar q\to \tchi^0_i \tilde g~~,~~ gg\to \tchi^0_i \tilde g~~,~~
g q\to \tchi^0_i \tilde q_{L,R}~~,~~
q\bar q'\to \tchi^0_i \tchi^{\pm}_j~~,~~
\label{processes}
\eq
where the indices $(i,j)$ now enumerate the neutralino and chargino respectively.
The aim of the present paper is  to study the physical consequences of these
subprocesses using  the same procedure as in \cite{pp-2chi0}. Since different particles
are involved in each of them, their  combined study is  sensitive to different aspects of the
underlying model.

\begin{figure}[t]
\[
\hspace{-0.5cm}\epsfig{file=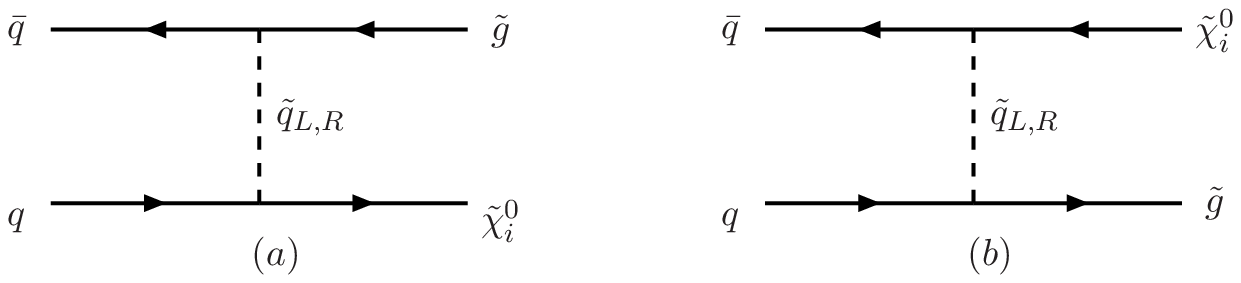,height=2.8cm, width=11cm}
\]
\caption[1]{Born  diagrams for  $q \bar q \to \tchi^0_i \tilde g$ }
\label{qq-gluino-fig}
\end{figure}

For the first, third and fourth of the subprocesses in (\ref{processes}), this model sensitivity
arises already at the Born level mainly  caused by the
(gaugino-higgsino) mixing matrices
\footnote{The notation of \cite{Rosiek} is used here.}
multiplying  the basic gaugino and higgsino couplings.
The relevant diagrams are shown
in Figs.\ref{qq-gluino-fig}-\ref{qq-chargino-fig} .
This model sensitivity is further enhanced when including also
the leading logarithmic part of the 1-loop corrections, calculated by
following  the procedure of \cite{BRV}.
Thus, rather simple expressions for the amplitudes
of these processes are  reached, which apart
from being very sensitive to the physical dynamics,
should also be quite adequate for LHC energies and accuracies.

Further model sensitivity  is induced in the case of $\tchi^0_i \tilde g$ production,
 by the contribution of the  genuine 1-loop subprocess $gg\to \tchi^0_i \tilde g$.
The generic form of the relevant diagrams is shown in Fig.\ref{gg-gluino-fig}.
On the basis of these, a numerical Fortran  code called PLATONgluino
is released, calculating $\d\sigma(gg\to \tchi^0_i \tilde g)/d\hat t $
for any set of real $\mu$ and MSSM soft breaking parameters
at the electroweak scale \cite{Platon}.

To explore the actual physical situation that might be realized within the  SUSY approach,
 typical MSSM benchmark models with real parameters are used \cite{Snowmass, Arnowitt, CDG}.
The LHC cross sections for proton proton collisions are then computed
by convoluting the $q\bar q$, $gg$, $qg$
subprocess cross sections,
with the corresponding quark
and gluon distribution functions taken from \cite{MRST} . As in \cite{pp-2chi0},
 invariant mass  and angular distributions are constructed,
 illustrations of which are given below.

The results obtained in this paper should be useful for precise
applications at LHC taking into account decay branching ratios
and final state identifications. We will come back to this point
in the conclusion.

The organization of the paper is the following. In Section 2.1 and Appendix A.1, the
general form of the  Born amplitudes for $q\bar q\to \tchi^0_i \tilde g $ are given,
together with the 1-loop  LL  EW
and\footnote{The  SUSY-QCD corrections describe a
special  part of the complete QCD correction,  intimately related  to
the SUSY dynamics. The fact that we consider them together with the EW corrections,
rather than the pure QCD ones, is a matter of choice.
See \cite{qqLHC} for its exact definition.} SUSY QCD corrections to them,
as well as the explicit Born expressions for the helicity amplitudes.
The corresponding results  for $q g \to \chi^0_i \tilde q_{L,R}$ and
$q\bar q'\to \tchi^0_i \tchi^{\pm}_j$ are given in Sections 2.2 and 2.3, and Appendices
A.2  and A.3 respectively; while in  Section 3, the 1-loop process
 $gg\to \tchi^0_i \tilde g$ is discussed. Finally, in Section 4 we discuss our results,
 and Section 5 presents  the Conclusions.\\

\section{The processes  $q\bar q\to \tchi^0_i \tilde g$,~
$g q\to \tchi^0_i \tilde q_{L,R}$,~
$q\bar q'\to \tchi^0_i \tchi^{\pm}_j$.}

The momenta, energies and masses in  these subprocesses,
as well as in  $gg \to \tchi^0_i \tilde g$ of Section 3,
are defined as
\bq
a(q_1)~ b(q_2)~ \to ~A(p_1, E_1, m_i)~ B(p_2, E_2, m_j)~~,  \label{prosses-kinematics}
\eq
where the  masses of the incoming particles are neglected.
Denoting by  $(p,~\theta)$  the final state c.m. momentum
and scattering angle, we have
\bqa
&& \s= (p_1+p_2)^2=(q_1+q_2)^2 ~~,  \nonumber \\
&& \t= (q_2-p_2)^2=(p_1-q_1)^2  ~~, \nonumber \\
&& \u= (q_2-p_1)^2=(p_2-q_1)^2  ~~,  \nonumber \\
&& p=\frac{1}{2\sqrt{\s}} \Big\{ [\s-(m_i+m_j)^2][\s-(m_i-m_j)^2] \Big\}^{1/2}~~,
\nonumber \\
&& \beta=\frac{2 p}{\sqrt{\s}} ~~~, ~~
   E_1=\frac{\s +\mid-\mjd}{2\sqrt{\s}}~~~,~~~
 E_2=\frac{\s +\mjd-\mid}{2\sqrt{\s}}~~~, \nonumber \\
&& q_1=\frac{\sqrt{\s}}{2}(1, 0,0,1) ~~~,~~~  q_2= \frac{\sqrt{\s}}{2}(1, 0,0,-1) ~, \nonumber \\
&& p_1= (E_1, p \sin\theta, 0, p \cos\theta) ~~,~~
p_2= (E_2, -p \sin\theta, 0, -p \cos\theta)  ~~. \label{kinematics}
\eqa

The common characteristic of the  subprocesses of the present Section,
is that they all receive
non-vanishing Born contributions determined by the diagrams in
Figs.\ref{qq-gluino-fig}-\ref{qq-chargino-fig}.
Since we neglect initial masses,
the only needed vertices for calculating the diagrams for  the first two processes,
are those given by  the
neutral gaugino-quark-squark couplings
\bq
A^{0L}_i(\tilde{u}_L)=-~{e\over 3\sqrt{2}s_Wc_W}
(Z^N_{1i}s_W+3Z^N_{2i}c_W)~~,~~
A^{0L}_i(\tilde{d}_L)=-~{e\over 3\sqrt{2}s_Wc_W}
(Z^N_{1i}s_W-3Z^N_{2i}c_W)~,\label{chi0-Lsquarks}
\eq
\bq
A^{0R}_i(\tilde{u}_R)=~{2e\sqrt{2}\over 3c_W}
Z^{N*}_{1i}~~,~~
A^{0R}_i(\tilde{d}_R)=-~{e\sqrt{2}\over 3c_W}
Z^{N*}_{1i}~, \label{chi0-Rsquarks}
\eq
and the corresponding chargino-$q\tilde q'_{L,R}$-ones
\bq
A^{cL}_j(\tilde u_L)=-\frac{e}{\sw}Z^+_{1j}~~,~~
A^{cL}_j(\tilde d_L)=-\frac{e}{\sw}Z^-_{1j}~~.
\label{chip-Lsquarks}
\eq
The notation of \cite{Rosiek} is used for the neutralino and chargino mixing matrices, and
$i$ and $j$ in (\ref{chi0-Lsquarks}, \ref{chi0-Rsquarks}) and (\ref{chip-Lsquarks}),
 denote the neutralino and chargino index respectively.\par
For the third process $q\bar q'\to \tchi^0_i \tchi^{\pm}_j$,
determined by  the three Born diagrams of Figs.\ref{qq-chargino-fig}a,b,c
one needs in addition the $W$-chargino-neutralino
couplings\footnote{We use the same notation as
in \cite{pp-2chi0, Rosiek}.}
\bq
O^{WL}_{ji}=Z^N_{2i}Z^{+*}_{1j}-{1\over\sqrt{2}}Z^N_{4i}Z^{+*}_{2j}
~~~,~~~ O^{WR}_{ji}=Z^{N*}_{2i}Z^{-}_{1j}+{1\over\sqrt{2}}Z^{N*}_{3i}Z^{-}_{2j}~~.
\label{OWLRji}
\eq

\subsection{The process $q\bar q\to \tchi^0_i \tilde g$ to the LL 1-loop EW order.}

Writing this process in more detail as
\bq
q_{c_1}(q_1,\lambda_1)~ \bar q_{ c_2}(q_2,\lambda_2)~
\to ~\tchi^0_i(p_1,\tau_1)~ \tilde g_l(p_2, \tau_2)~~,
 \label{process-gluino}
\eq
we denote by   $(c_1, c_2)$  the color indices
for $(q,~\bar q)$ respectively, and by $l$  the color index of
$\tilde g$. The helicity amplitude  is then written as
$F_{\lambda_1\lambda_2;\tau_1\tau_2}$,
with color indices suppressed and $\lambda_1,\lambda_2,\tau_1,\tau_2$
denoting the helicities. The mass $\mj$ in (\ref{kinematics}) now describes
 the gluino mass.

The  Born level contributions to this amplitude arising from  the two diagrams,
in Fig.\ref{qq-gluino-fig}a,b are
\bqa
F^{\rm B (a)}&=&- {g_s\sqrt{2}A^{0L}_i(\tilde q_L)
\over \t-m^2_{\tilde q_L}} \Big ({\lambda^l\over2}\Big )_{c_2c_1}
[\bar v(q_2)P_Ru^c(\tilde g)]
[\bar u(\chi^0_i)P_Lu(q_1)]\nonumber\\
&& + {g_s\sqrt{2}A^{0R}_i(\tilde q_R)
\over \t-m^2_{\tilde q_R}} \Big ({\lambda^l\over2}\Big )_{c_2c_1}
[\bar v( q_2)P_Lu^c(\tilde g)] [\bar u(\chi^0_i)P_Ru(q_1)] ~~, \nonumber \\
F^{\rm B (b)}&=& {g_s\sqrt{2}A^{0L*}_i(\tilde q_L)
\over \u-m^2_{\tilde q_L}}
\Big ({\lambda^l\over2}\Big )_{c_2c_1}
[\bar v(q_2)P_Ru^c(\chi^0_i)]
[\bar u(\tilde g)P_Lu(q_1)]\nonumber\\
&&- {g_s\sqrt{2}A^{0R*}_i(\tilde q_R)
\over \u-m^2_{\tilde q_R}}
\Big ({\lambda^l\over2}\Big )_{c_2c_1}
[\bar v(q_2)P_Lu^c(\chi^0_i)]
[\bar u(\tilde g)P_Ru(q_1)]~. \label{Born-amp-gluino}
\eqa
\noindent
where (\ref{chi0-Lsquarks},\ref{chi0-Rsquarks})
have been used and $g_s$ denotes the QCD coupling.

The explicit expressions of the Born  helicity amplitudes
$F^B_{\lambda_1\lambda_2,\tau_1\tau_2}$
are given in (\ref{Born-amp-gluino-app}). To get full helicity amplitudes containing  also
 the 1-loop  LL  EW and SUSY QCD contributions, the corrections
 in (\ref{universal-amp-gluino-cor})
and (\ref{angular-amp-gluino-cor}) should be added.
 The differential cross section is then obtained as
\bq
{d\sigma (q\bar q\to \tchi^0_i \tilde g)
\over d \cos\theta}={\beta \over 1152 \pi s}
\sum_{\rm col, ~spins}|F_{\lambda_1\lambda_2,\tau_1\tau_2}|^2 ~~.
\label{dsigma-qq-gluino}
\eq

At asymptotic energies, much larger than all masses,
both the dominant amplitudes (see (\ref{qq-gluino-amp-asym})), and the
differential cross sections simplify considerably. \\

\subsection{The process $ q g\to \tchi^0_i \tilde q_{L,R}$ to the LL
1-loop EW  order}

Writing this  process as
\bq
q_{c_1}(q_1,\lambda_1)  g_l(q_2,\mu_2)
\to \tchi^0_i(p_1,\tau_1) \tilde q_{Ic_2}(p_2)~~, \label{prosses-squark}
\eq
 we denote by  $(c_1,~c_2, ~l)$  the color indices
for $(q,~\tilde q_I,~ g )$ respectively, while  $(I=L,~R) $ determines the
type of the produced squark of the first or second family.
The helicities of initial quark and gluon, as well as the
helicity of the final neutralino, are
respectively described by $\lambda_1, ~\mu_2, ~\tau_1$. Correspondingly, the polarization
vector of the initial gluon is  denoted as $\epsilon_2$, and the
full helicity amplitudes for the  process is written  as
$F_{\lambda_1 \mu_2;\tau_1}$.
As before, the kinematics are fixed by (\ref{prosses-kinematics}, \ref{kinematics}),
with  $\mj$ now describing the squark mass.

\begin{figure}[b]
\[
\hspace{-0.5cm}\epsfig{file=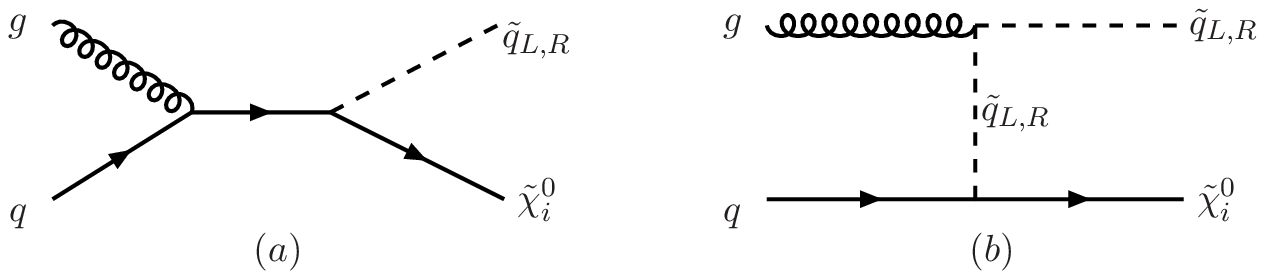,height=2.8cm, width=11cm}
\]
\caption[1]{Born diagrams for  $q g \to \tchi^0_i  \tilde q_{(L,R)} $. }
\label{qg-squark-fig}
\end{figure}

The two relevant Born diagrams are  shown  in Fig.\ref{qg-squark-fig}.
 Since the incoming quark is massless, the squark specification by the index
$(I=L,~R) $,  is uniquely  associated with the quark helicity  being $( \lambda_1=-1/2, ~+1/2)$
respectively; this property remaining true  at 1-loop LL level also.

With the momenta and helicities defined  by (\ref{prosses-squark}), the contributions
from the diagrams in Figs.\ref{qg-squark-fig}a,b may then be written as
\bqa
F^{\rm B(a)} &= & {g_sA^{0I}_i(\tilde q_I )\over \s}
\Big ({\lambda^l\over2}\Big )_{c_2c_1}
[\bar u(\chi^0_i)P_I (\rlap/ q_1+\rlap / q_2)) \rlap / \epsilon_2 u(q_1)] ~~, \nonumber \\
F^{\rm B(b)}& = &  {g_sA^{0I}_i(\tilde q_I)\over \t-m^2_{\tilde q_I} }
\Big ({\lambda^l\over2}\Big )_{c_2c_1}
[\bar u(\chi^0_i)P_I u(q)](2\epsilon_2 .p_2 )~~.\label{Born-amp-squark}
\eqa
The resulting Born helicity amplitudes appear in
(\ref{Born-amp-squark1-app}, \ref{Born-amp-squark2-app}), while their asymptotic expressions
are given in (\ref{Born-amp-asym-squark2}). The universal and angular parts of the
LL EW and SUSY corrections to these amplitudes are respectively given  in
(\ref{universal-amp-cor-squark}, \ref{angular-amp-cor-squark}).

After averaging over spins and colors, the cross sections are obtained from these amplitudes
by
\bq
{d\sigma(q g\to \tchi^0_i \tilde q_I)\over d \cos\theta}
=  {\beta\over3072\pi s} \sum_{col,spins}|F_{\lambda_1,\mu_2,\tau_1}|^2 ~~.
\label{dsigma-qg-squark}
\eq

The asymptotic expressions of the amplitudes including all LL EW and SUSY QCD corrections
appear  in (\ref{amp-asym-squark}).  \\

\subsection{The process $q\bar q'\to \tchi^0_i \tchi^\pm_j$
to the LL 1-loop EW  order.}

The two contributing  processes in this case, namely $u\bar d \to \tchi^0_i+\tchi^+_j$ and
$d\bar u \to \tchi^0_i+\tchi^-_j$,  should give equal differential cross sections,
because of the CP invariance valid for real soft MSSM breaking and $\mu$ parameters:
\bq
\frac{\d\sigma (u\bar d \to \tchi^0_i+\tchi^+_j)}{\d\cos\theta}=
\frac{\d\sigma (d\bar u \to \tchi^0_i+\tchi^-_j)}{\d\cos\theta}~~.  \label{CP-chargino}
\eq

We therefore concentrate on
\bq
u(q_1,\lambda_1)  \bar d(q_2,\lambda_2)~ \to \tchi^0_i(p_1,\tau_i) ~ \tchi^+_j(p_2, \tau_j)~~,
 \label{prosses-chi0p}
\eq
where the helicities and momenta are defined, so that (\ref{kinematics}) keeps describing the
kinematics with $\mj $ now being the chargino mass.
The helicity amplitudes are denoted as
$F_{\lambda_1 \lambda_2;\tau_i\tau_j}$.

 The Born level  contributions arise
 from the three  diagrams in Figs.\ref{qq-chargino-fig}a,b,c  caused  respectively
 by the exchanges of a $W^+$ in the s-channel, a $\tilde u_L$-squark in the $t$-channel, and
a $\tilde d_L$-squark in the $u$-channel, and suitably analyzed as
\bq
F^{ijB}_{\lambda_1 \lambda_2;\tau_i\tau_j}=
S^{ijB}_{\lambda_1 \lambda_2;\tau_i\tau_j}
+T^{ijB}_{\lambda_1 \lambda_2;\tau_i\tau_j}
+U^{ijB}_{\lambda_1 \lambda_2;\tau_i\tau_j} ~~ ,\label{Born-amp-chi0p}
\eq
where the indices $(i,j )$ refer to the neutralino and chargino  respectively.
Note that, since we neglect  quark masses, there are  no R-squark exchange
contributions.

Defining  the momenta and helicities as  in (\ref{prosses-chi0p}), and
using (\ref{OWLRji}),
the   contributions from the three diagrams in  Figs.\ref{qq-chargino-fig}a,b,c
to  the Born helicity amplitudes appear in (\ref{chi0p-Born-s}, \ref{chi0p-Born-t},
\ref{chi0p-Born-u}) respectively.

At asymptotic energies,  only $F^{ij}_{-+-+}$ and $F^{ij}_{-++-}$ retain a  non-vanishing
 Born contribution appearing in (\ref{chi0p-Born-asym1},\ref{chi0p-Born-asym2});
 while the associated EW universal, SUSY QCD, RG and angular LL corrections are
shown respectively in (\ref{Univ-chi0p-mpmp}, \ref{Univ-chi0p-mppm}),
(\ref{SQCD-chi0p}), (\ref{RG-chi0p}) and (\ref{ang-chi0p-mpmp},\ref{ang-chi0p-mppm}).

The spin and color averaged differential cross section is calculated from
\bq
{d\sigma (u \bar d\to \tchi^0_i \tchi^+_j)\over d \cos\theta}
={d\sigma (d \bar u\to \tchi^0_i \tchi^-_j)\over d \cos\theta}
={\beta \over 384 \pi s}
\sum_{\rm  spins}|F^{ij}_{\lambda_1\lambda_2,\tau_1\tau_2}|^2 ~~. \label{dsigma-chi0p}
\eq\\

\section{The one loop process $gg\to \tchi^0_i \tilde{g}$ }

The momenta,  helicities and color indices $(a_1,a_2,a_3)$ of the particles participating
in this process, together with  the polarization vectors of the gluons,
are defined though
\bq
g_{a_1}(q_1, \epsilon_1(\mu_1))+ g_{a_2}(q_2, \epsilon_2(\mu_2))
\to
\tchi^0_i(p_1, \lambda_1)~+~\tilde g_{a_3}(p_2, \lambda_2)
 ~~ . \label{process-gg-gluino}
\eq
The kinematics is defined in (\ref{kinematics}), with $\mi$ denoting the neutralino mass and
$\mj$ the mass of the gluino. The helicity amplitude of the process  denoted as
$F_{\mu_1\mu_2; \lambda_1\lambda_2}^{a_1a_2; a3}(\theta)$,
 satisfies
\bq
F_{\mu_1\mu_2; \lambda_1\lambda_2}^{a_1a_2; a3}(\theta)
=(-1)^{\lambda_1-\lambda_2} F_{\mu_2\mu_1; \lambda_1\lambda_2}^{a_2a_1; a3}(\pi-\theta)
~~, \label{gg-Bose}
\eq
because of   Bose symmetry among the initial  gluons.

This process first appears at the 1-loop level, driven by the diagrams generically shown in
Fig.\ref{gg-gluino-fig}. These consist of three types of box diagrams named (B1, B2, B3),
which are of exactly the same form as those met in neutralino pair production
in an $LC_{\gamma \gamma}$ collider, or in the calculation of
the reverse  process of dark matter annihilation
to photons \cite{gamma-gamma, DMann}. In addition to them, there are
three types of s-channel triangular diagrams
(s1, s2, s3), and two types of t-channel triangles.
These diagrams have been calculated exactly and the  results were
 used to construct the  FORTRAN code named PLATONgluino which,
after averaging  over all spins and colors, calculates
\[
\frac{\d\sigma (g g \to \tchi^0_i \tilde g) }{\d\t} ~~~{\rm in } ~~~ fb/TeV^2
\]
for any value of the subprocess  c.m. scattering angle $\theta$
given in radians, and any set of real MSSM parameters at the electroweak scale. As with
other related codes we have constructed, PLATONgluino may be obtained from
\cite{Platon}.

\section{Results}

In this Section we discuss the LHC predictions for the direct production of
a single neutralino  associated with  a gluino, squark or  chargino,
according to the four subprocesses presented
in Sections 2 and 3. The predictions are valid for  any
MSSM model with real SUSY parameters. An   exploration
of the possible results has been made, using  typical benchmark models
\cite{Snowmass, Arnowitt, CDG}. These  benchmarks have also
been used in other recent neutralino explorations,
and their sole purpose is to help  identifying   the physical parameters
mainly affecting the neutralino production at LHC  \cite{DMann, gamma-gamma, pp-2chi0}.

For the parton  distribution functions inside
the proton, we use  the MRST2003c package \cite{MRST}
at the factorization scale
\bq
Q=\frac{E_{Ti}+E_{Tj}}{4}~~. \label{Q-scale}
\eq
A complete summary of  the relevant parton formulae  and  kinematics
may be found in  Appendix B of \cite{pp-2chi0}.

As observables we use  the invariant mass distribution
$d\sigma/d\s$ of the aforementioned subprocesses, and the
  c.m. angular distribution  $d\sigma/d\chi$ defined \eg in
eqs.(B.39,B.43) of \cite{pp-2chi0}. The  $\chi$-variable  is always taken
to describe  the particle accompanying the neutralino
in the subprocess and is defined in terms of c.m. variables
\bqa
\chi_j\equiv e^{2y_j^*}&= & {1 - {p^*\over E^*_j} \cos\theta^*
\over 1 +{p^*\over E^*_j} \cos\theta^*} ~~ , \label{chi} 
\eqa
\noindent
so that our treatment covers both, the LSP $\tchi^0_1$ case, as well
as the case of a heavier $\tchi^0_i$.
The transverse momentum distribution is not shown in any detail here,
since it presents  the same features as the mass distribution;
a similar situation has already been noticed for $\tchi^0_i\tchi^0_j$ production
  \cite{pp-2chi0}. Depending on the experiment of course,
 such  distributions may also be useful. \\

Extensive  sensitivity of the single neutralino production processes
to the SUSY MSSM parameters, is observed.
This is caused mainly by  the dependence of  the neutralino couplings
 on the percentage of their gaugino (Bino and Wino)
or higgsino components, through the $Z^N_{ji}$ mixing matrices
\cite{Rosiek}. The four processes in (\ref{processes}) react differently to
this percentage, as the first three  are  mainly controlled
by the gaugino components\footnote{Some dependence on the higgsino component appears
also for the  subprocess $gg\to \tchi^0_i \tilde g$, caused by
$(\tilde t, \tilde b)$-loop contributions. It should be remembered though that
the contribution of this subprocess, being of higher order,
is generally suppressed.}, whereas the fourth process
depends both, on the gaugino and on the higgsino
components.

As we are especially interested in the structure of the
LSP, supposedly the lightest $\tchi^0_1$, which, depending on
the benchmark, can predominantly be either Bino, or Wino, or
higgsino, this explains the large sensitivity to the
chosen benchmark.\par

 The single neutralino production processes
are also very  sensitive to the masses  of the exchanged squarks.
In the gluino  case, the relative importance of the one loop process $gg\to\tchi^0_i\tilde{g}$,
is also strongly depending  on the  squark masses.
For light squarks, this  gives cross sections which are about
a hundred times smaller than the ones from the $q\bar q$ process.
But if the squarks of the first two generations  become  heavy, while those of the third
remain rather light, it may turn out that kinematical regions exist where the
1-loop subprocess $gg\to\tchi^0_i\tilde{g}$ is  appreciable, compared to the
Born-level subprocess $q \bar q \to\tchi^0_i\tilde{g}$; so that it
cannot be ignored. Comparing with the   $gg\to\tchi^0_i\tchi^0_k$ treatment of \cite{pp-2chi0},
we should note  that  the $gg\to\tchi^0_i \tilde{g}$ process cannot be enhanced
by  resonance effects like those enhancing  $\tchi^0_i\tchi^0_k$ production. \par

For what concerns the electroweak radiative corrections to the three
Born processes, the computations of the leading and subleading
logarithmic contributions show a
negative effect, regularly increasing with the invariant mass,
which is of the order of (10-20)\% at the TeV range, as expected from \cite{BRV}.
This effect is comparable, but of opposite sign, to the analogous
QCD correction\cite{Spiratalk,Beenakker,Plehn},
and  it should also be taken into account
for a precise analysis.\\

 On the  basis of our explorations, we present
 in Figs.\ref{SPS1a-mass-fig}-\ref{ADfg9a-mass-chi-fig} below,
 illustrations for  four different
typical cases\footnote{In all cases we have used  ${\rm Suspect2\_3}$  to calculate
the various masses \cite{suspect}.}:\\
\noindent
1) a gaugino (Bino)-type model for $\tchi^0_1$, with light squark masses, SPS1a
\cite{Snowmass},\\
1a) a Bino-type model with heavy squark masses\footnote{SPS1aa is constructed
from SPS1a of \cite{Snowmass}, by simply putting the high scale SUSY breaking soft
sfermion masses of   the first and second generations, at 5000 GeV.},  SPS1aa, \\
2) a higgsino type model for $\tchi^0_1,~\tchi^\pm_1$,
with light squark masses\footnote{This model is
extracted from Fig.9 of \cite{Arnowitt}. It is characterized by the high
scale values: $m_{1/2}=420 ~GeV$,
$A_0=420~ GeV$, $\tan\beta=40$, and $m_0=600~ GeV$ for all scalar masses except
$m_{H_u}=600 \sqrt{2} ~GeV$.   To preserve the predominantly higgsino nature of $\tchi^0_1$,
 $m_t=174 ~GeV$ should be used here, as this was the case when the model was constructed.},
 AD(fg9) \cite{Arnowitt},\\
2a) a related higgsino type model, but
 with heavy squark masses\footnote{It is constructed from the EW
scale masses of AD(fg9), by only changing
the sfermion soft SUSY breaking masses of the first two
generations, which are now put at 5000 GeV.},  AD(fg9a) \cite{Arnowitt},\\
\noindent
discussed in turn below:

\begin{itemize}

\item   Let us first consider
the $\tchi^0_1$ gaugino-type model (SPS1a). This model  gives invariant mass distributions
for the three Born processes in (\ref{processes}),
which are  largely observable in the 1 TeV range; \ie  cross sections of
about 100fb for the first two cases, but only 10fb or less
for the third one. It also gives an  invariant mass distribution for  the
$ q g\to \tchi^0_i \tilde q_{L,R}$ channel, which
is more important at low masses,  due to the behavior of the gluon
distribution function. The angular distribution
for $q\bar q \to \tchi^0 \tilde g$,  described  at Born
level  by  the t- and u- channel squark exchanges indicated in
Fig.\ref{qq-gluino-fig},    flattens out
at large $\chi_{\tilde g}$ in this model.

It is amusing to remark that a very similar behavior is also  expected in the
universal m-SUGRA type model which has been identified by
\cite{Ellis}, as "a best"  description of all
present particle and cosmological constraints \cite{DMpresent}.

\item  Comparing  SPS1aa  to SPS1a, one notices
a reduction of the invariant mass distribution in
the 1 TeV range, by more than an order of magnitude for
$\tchi^0_i \tilde g$,
by somewhat less than an order of magnitude for  $\tchi^0_i\tchi^\pm_1$;
and, obviously, a complete suppression of the $\tchi^0_i \tilde q$ production.
Moving from SPS1a to SPS1aa, there appears also a change in
the $\chi$-distribution, which becomes  steeper for  $\tchi^0_i \tilde g$,
but  remains  roughly similar for  $\tchi^0_i\tchi^\pm_1$.

\item  We next turn to  the higgsino type model AD(fg9)
for the three channels studied here. Comparing them to the gaugino SPS1a and SPS1aa
 models, we find that the predicted cross sections for
the  $\tchi^0_i \tilde g$ and $\tchi^0_i \tilde q$ channels, are much smaller now,
since they  are controlled essentially  by the
gaugino components; (see Figs.\ref{ADfg9-mass-fig},\ref{ADfg9a-mass-chi-fig}).
On the contrary, the cross section for the $\tchi^0_i\tchi^\pm_1$ process
is   larger,  because of the presence of the
s-channel W exchange  diagram involving   higgsino
components; (see Fig.\ref{qq-chargino-fig}). The $\chi$-distribution
becomes also flatter, for the same reason. These can be seen by
comparing Figs.\ref{ADfg9-mass-fig}c,\ref{ADfg9-chi-fig}c and \ref{ADfg9a-mass-chi-fig}b,d
with Figs.\ref{SPS1a-mass-fig}c,\ref{SPS1a-chi-fig}c and \ref{SPS1aa-mass-chi-fig}b,d
respectively.

\item  Finally, we  compare the results of the models
AD(fg9) and AD(fg9a), in both  of which there is a large
 higgsino component to $\tchi^0_1$, as well as to  $\tchi^0_2$.
In AD(fg9a) the squark contribution, coupled through the
gaugino component, is further reduced compared to AD(fg9), leading
to an even smaller  prediction
for the $\tchi^0_i \tilde g$; compare  Figs.\ref{ADfg9-mass-fig}a,\ref{ADfg9-chi-fig}a
with \ref{ADfg9a-mass-chi-fig}a,c. On the other hand,
for the  $\tchi^0_1\tchi^\pm_1$ and $\tchi^0_2\tchi^\pm_1$ channels,  the cross
sections are comparable, since they both receive a large contribution from
the higgsino component coupled to the intermediate $W$ boson,
which is not affected by the change in the squark mass;
compare  Figs.\ref{ADfg9-mass-fig}c,\ref{ADfg9-chi-fig}c
with \ref{ADfg9a-mass-chi-fig}b,d.

\end{itemize}

Finally we   comment on the difference between the  magnitude of
 the $\tchi^0_1$ production cross section (in which we are mainly
interested in),  and the  $\tchi^0_2$ one;
$\tchi^0_2$ production is generally more copious than the
$\tchi^0_1$ one,  becoming   progressively more pronounced
as we go from  the $\tchi^0_i \tilde g$ channel, to $\tchi^0_i \tilde q$, and eventually to
the $\tchi^0_i\tchi^\pm_1$ channel, where it reaches a factor 10 in the SPS1a model.
This factor is even larger in the SPS1aa model; of order 100.
These differences are due to the
$Z_{ji}$ mixing matrix elements appearing in the Born amplitudes, which control  the
Bino, Wino and higgsino components of the neutralinos.\\

\section{General Conclusion on $\tchi^0_i $ production}

In this paper we have analyzed the  single
neutralino production processes
$\tchi^0_i \tilde g$, $\tchi^0_i \tilde g$ and
$\tchi^0_i \tchi^{\pm}_j$ at LHC.\par

The complete set of helicity amplitudes
for the Born terms of the subprocesses
$q\bar q\to \tchi^0_i \tilde g$, $g q\to \tchi^0_i \tilde q_{L,R}$,
$q\bar q'\to \tchi^0_i \tchi^{\pm}_j$ has been written down,
together with  the leading and subleading logarithmic electroweak
corrections to them. Compact analytic expressions are presented,
which are applicable to any MSSM model with real parameters.
We have also included  the complete  one loop calculation of the
subprocess $gg\to \tchi^0_i \tilde g$, for which a numerical code called PLATONgluino
is released \cite{Platon}.\par

 The pure QCD corrections, which have  already been
given in previous papers \cite{Beenakker, Plehn},  have not been reexamined.
But  we have emphasized  that, contrary may be to naive expectations,
the  leading  logarithmic EW and QCD corrections at the LHC energies have
similar  magnitudes, but  opposite sign.
Thus, they should both be taken into account in analyzing the experimental
 data. \par

The single neutralino production processes have been found to be mainly sensitive on
two physical sets of quantities; namely  the amount of gaugino and higgsino components
of the neutralinos, and the scale of the soft breaking parameters for the
squarks of the first and second generations. To emphasize this, a set of
illustrations for LHC invariant mass and angular distributions have been presented
which indeed show this sensitivity. These  were based on
four "benchmark" models, but  more were explored in our actual runnings.\par

This physics output should of course be joint to the one that can be obtained
from the $\tchi^0_i \tchi^0_j$ production  studied previously \cite{pp-2chi0}.
For that purpose we have added
Figs.\ref{SPS1a-chi0chi0-fig},\ref{SPS1aa-chi0chi0-fig}, which show the
 invariant mass and  $\chi$ distributions for
$\tchi^0_2 \tchi^0_1$ and $\tchi^0_2 \tchi^0_2$  production, in the same
SPS1a and the SPS1aa models used  for the single neutralino
case\footnote{Analogous results have been shown in Figs.11,12 of \cite{pp-2chi0},
where though, $\chi$ is defined as the inverse of  the present one.}.
In going from SPS1a to SPS1aa, one sees a
reduction of the Born contribution, rather similar to what happens
in the $\chi^0_{1,2}\tilde{g}$ case; but one also sees that the
relative role of the one loop $gg$ process in SPS1aa, is more important for
 $\tchi^0_2 \tchi^0_1$  production, then for $\chi^0_{1,2}\tilde{g}$.
So the neutralino pair production  channel has its own
typical features.\par

Summarizing, we have observed that the channels
$\tchi^0_i \tilde g$, $ \tchi^0_i \tilde g$,
$\tchi^0_i \tchi^{\pm}_j$ and   $\tchi^0_i \tchi^0_j$
present an important sensitivity to the neutralino  structure;
particularly to  the relative magnitude  of its gaugino
and higgsino components.
They also present a considerable
sensitivity to  the MSSM mass spectrum
for the  gluino, squarks, charginos and Higgses;
the later being able to lead
to possible resonance effects.\par

We conclude by emphasizing that the results obtained
in this paper should  be completed by detail
experimental studies dedicated for LHC.  Observables should then be
constructed addressing neutralino, gluino and squark
decay channels to   various  numbers of  jets and leptons.
Such  observables should also
reflect, at some important level, the sensitivity to the neutralino
properties\footnote{If discovered at LHC or
a Linear Collider, it would also be the first time
that physical quantities highly sensitive to the Majorana
nature of a particle reach such a high   observability.}, that we have
observed at the level of the basic processes.

We hope that their measurement will be able to
confirm or infirm the possibility
that the neutralino is an important component
of the Dark Matter of the Universe.\\

\noindent
{\large\bf{Acknowledgement}}\\
\noindent
G.J.G. gratefully acknowledges also the support by the European Union  RTN contract
MRTN-CT-2004-503369, and the hospitality extended to him by the CERN Theory
Division during the later  part of this work.

\newpage
\appendix

\renewcommand{\thesection}{Appendix A.\arabic{section}}
\renewcommand{\theequation}{A.\arabic{equation}}
\setcounter{equation}{0}

\section{ Helicity amplitudes for $q\bar q\to \tchi^0_i \tilde g $
\label{gluino-appendix}}

Starting from (\ref{Born-amp-gluino}), the explicit expressions of the
Born helicity amplitudes for the process shown in (\ref{process-gluino}),   are
\bqa
 F^B_{-+++}&= &\tilde C_{\tg} \sin \theta
 \Bigg \{ -\frac{A^{0L}_i(\tilde q_L)}{\t-m^2_{\tilde q_L}}
 \Big [(m_i+m_{\tg})\sqrt{ \s-(m_i-m_{\tg})^2} -(m_i- m_{\tg}) \sqrt{ \s-(m_i+m_{\tg})^2}\Big ]
 \nonumber \\
 & + &\frac{A^{0L*}_i(\tilde q_L)}{\u-m^2_{\tilde q_L}}
 \Big [(m_i+m_{\tg})\sqrt{ \s-(m_i-m_{\tg})^2} +(m_i- m_{\tg}) \sqrt{ \s-(m_i+m_{\tg})^2}\Big ]
 \Bigg \}
 ~~, \nonumber \\
 F^B_{+-++}&= &\tilde C_{\tg} \sin \theta
 \Bigg \{ -\frac{A^{0R}_i(\tilde q_R)}{\t-m^2_{\tilde q_R}}
 \Big [(m_i+m_{\tg})\sqrt{ \s-(m_i-m_{\tg})^2} +(m_i- m_{\tg}) \sqrt{ \s-(m_i+m_{\tg})^2} \Big ]
 \nonumber \\ & + &\frac{A^{0R*}_i(\tilde q_R)}{\u-m^2_{\tilde q_R}}
 \Big [(m_i+m_{\tg})\sqrt{ \s-(m_i-m_{\tg})^2} -(m_i- m_{\tg}) \sqrt{ \s-(m_i+m_{\tg})^2}\Big ]
 \Bigg \} ~~, \nonumber \\
 F^B_{-+--}&= &\tilde C_{\tg} \sin \theta
 \Bigg \{ -\frac{A^{0L}_i(\tilde q_L)}{\t-m^2_{\tilde q_L}}
 \Big [(m_i+m_{\tg})\sqrt{ \s-(m_i-m_{\tg})^2} +(m_i- m_{\tg}) \sqrt{ \s-(m_i+m_{\tg})^2}\Big ]
 \nonumber \\
 & + &\frac{A^{0L*}_i(\tilde q_L)}{\u-m^2_{\tilde q_L}}
 \Big [(m_i+m_{\tg})\sqrt{ \s-(m_i-m_{\tg})^2} -(m_i- m_{\tg}) \sqrt{ \s-(m_i+m_{\tg})^2}\Big ]
 \Bigg \} ~~, \nonumber \\
 F^B_{+---}&= &\tilde C_{\tg} \sin \theta
 \Bigg \{ -\frac{A^{0R}_i(\tilde q_R)}{\t-m^2_{\tilde q_R}}
 \Big [(m_i+m_{\tg})\sqrt{ \s-(m_i-m_{\tg})^2} -(m_i- m_{\tg}) \sqrt{ \s-(m_i+m_{\tg})^2}\Big ]
 \nonumber \\
 & + &\frac{A^{0R*}_i(\tilde q_R)}{\u-m^2_{\tilde q_R}}
 \Big [(m_i+m_{\tg})\sqrt{ \s-(m_i-m_{\tg})^2} +(m_i- m_{\tg}) \sqrt{ \s-(m_i+m_{\tg})^2}\Big ]
 \Bigg \} ~~, \nonumber \\
 F^B_{-++-}&= &\tilde C_{\tg} (1-\cos \theta)
 \Bigg \{ -\frac{A^{0L}_i(\tilde q_L)}{\t-m^2_{\tilde q_L}} \sqrt{\s}
 \Big [ \sqrt{ \s-(m_i-m_{\tg})^2}
+\sqrt{\s-(m_i+m_{\tg})^2}\Big ]
\nonumber \\
& + &\frac{A^{0L*}_i(\tilde q_L)}{\u-m^2_{\tilde q_L}}
\sqrt{\s}\Big [ \sqrt{ \s-(m_i-m_{\tg})^2}
-\sqrt{\s-(m_i+m_{\tg})^2}\Big ]
\Bigg \} ~~, \nonumber \\
 F^B_{+--+}&= & - \tilde C_{\tg} (1-\cos \theta)
 \Bigg \{ -\frac{A^{0R}_i(\tilde q_R)}{\t-m^2_{\tilde q_R}} \sqrt{\s}
 \Big [ \sqrt{ \s-(m_i-m_{\tg})^2}
+\sqrt{\s-(m_i+m_{\tg})^2}\Big ]
\nonumber \\
& + &\frac{A^{0R*}_i(\tilde q_R)}{\u-m^2_{\tilde q_R}}
\sqrt{\s} \Big [ \sqrt{ \s-(m_i-m_{\tg})^2} -\sqrt{\s-(m_i+m_{\tg})^2}\Big ]
\Bigg \} ~~, \nonumber \\
 F^B_{-+-+}&= &\tilde C_{\tg} (1+\cos \theta)
 \Bigg \{ -\frac{A^{0L}_i(\tilde q_L)}{\t-m^2_{\tilde q_L}} \sqrt{\s}
 \Big [ \sqrt{ \s-(m_i-m_{\tg})^2}
-\sqrt{\s-(m_i+m_{\tg})^2}\Big ]
\nonumber \\
& + &\frac{A^{0L*}_i(\tilde q_L)}{\u-m^2_{\tilde q_L}}
\sqrt{\s} \Big [ \sqrt{ \s-(m_i-m_{\tg})^2}
+\sqrt{\s-(m_i+m_{\tg})^2}\Big ]
\Bigg \} ~~, \nonumber \\
 F^B_{+-+-}&= & - \tilde C_{\tg} (1+\cos \theta)
 \Bigg \{ -\frac{A^{0R}_i(\tilde q_R)}{\t-m^2_{\tilde q_R}} \sqrt{\s}
 \Big [ \sqrt{ \s-(m_i-m_{\tg})^2}
-\sqrt{\s-(m_i+m_{\tg})^2}\Big ]
\nonumber \\
& + &\frac{A^{0R*}_i(\tilde q_R)}{\u-m^2_{\tilde q_R}}
\sqrt{\s} \Big [ \sqrt{ \s-(m_i-m_{\tg})^2}
+\sqrt{\s-(m_i+m_{\tg})^2}\Big ]
\Bigg \} ~~, \nonumber \\
&& F^B_{++\tau \tau'}= F^B_{--\tau \tau'}=0~~, \label{Born-amp-gluino-app}
\eqa
where
\bq
\tilde C_{\tg}=-  \frac{g_s }{2 \sqrt{2}}
\Big ( \frac{\lambda^l}{2} \Big )_{ c_2  c_1} ~~, \label{Cgluino-alt}
\eq
with  $(c_1,~c_2, ~l)$ denoting   the color indices of the quark, antiquark and gluino
respectively, as defined in (\ref{process-gluino}). The kinematics is determined in
(\ref{kinematics}).

In the high energy limit, where  $ \s,~ |\t|,~ |\u|$ are all much larger
than all masses, the only non-vanishing Born  amplitudes simplify to
 \bqa
 F^{\rm B~asym}_{-++-} & \simeq &
- g_s\sqrt{2}A^{0L}_i(\tilde q_L) \Big
({\lambda^l\over2}\Big )_{ c_2  c_1} =2 e g_s
\Big [ \frac{Z^N_{1i} }{6 \cw} +\frac{I^{(3)}_q}{\sw} Z^N_{2i} \Big ]
\Big ({\lambda^l\over2}\Big )_{ c_2  c_1}~~, \nonumber \\ F^{\rm B~asym}_{+--+}
& \simeq &  g_s\sqrt{2}A^{0R}_i(\tilde q_R) \Big ({\lambda^l\over2}\Big )_{ c_2  c_1}=
2 e g_s \frac{Q_q}{\cw}  Z^{N*}_{1i}\Big
({\lambda^l\over2}\Big )_{ c_2  c_1}
~~, \nonumber \\
F^{\rm B~asym}_{-+-+} & \simeq &  g_s\sqrt{2}A^{0L*}_i(\tilde q_L)
\Big ({\lambda^l\over2}\Big )_{ c_2  c_1}
=- 2 e g_s \Big [ \frac{Z^{N*}_{1i}}{6 \cw}
+\frac{I^{(3)}_q}{\sw} Z^{N*}_{2i}
 \Big ] \Big ({\lambda^l\over2}\Big )_{ c_2  c_1} ~~, \nonumber \\
 F^{\rm B~asym}_{+-+-} & \simeq &
- g_s\sqrt{2}A^{0R*}_i(\tilde q_R) \Big
({\lambda^l\over2}\Big )_{ c_2  c_1}
=- 2 e g_s \frac{Q_q}{\cw}  Z^{N}_{1i}
\Big ({\lambda^l\over2}\Big )_{ c_2  c_1}
 ~~. \label{Born-amp-gluino-asym}
\eqa\\

The 1-loop universal leading logarithmic  EW and  SUSY QCD corrections only affect
these asymptotically dominant amplitudes and are given by \cite{BRV, Denner, qqLHC, Beenakker}
\bqa
F^{\rm Univ}_{-++-}&=&
F^{\rm B~asym}_{-++-} \cdot \R_{\tilde g L}
+(2I^3_q)c^{ew}(W)
{eg_s\over s_W}\Big ({\lambda^l\over2}\Big )_{ c_2  c_1} Z^{N}_{2i}~~,
\nonumber \\
F^{\rm Univ}_{-+-+}&=&
F^{\rm B~asym}_{-+-+} \cdot \R_{\tilde g L}
-(2I^3_q)c^{ew}(W)
{eg_s\over s_W}\Big ({\lambda^l\over2}\Big )_{ c_2  c_1} Z^{N*}_{2i}~~,
\nonumber \\
F^{\rm Univ}_{+--+}&=&
F^{\rm B~asym}_{+--+}\cdot \R_{\tilde g R}
~~,  \nonumber \\
F^{\rm Univ}_{+-+-}&=& F^{\rm B~asym}_{+-+-} \cdot \R_{\tilde g R}
~~, \label{universal-amp-gluino-cor}
\eqa
where $2I^3_q=\pm 1$, depending on whether $q=u$ or $d$.
 The universal LL correction due to the $q\bar q$-pair  is contained
in the parameter
\bq
\R_{\tilde g H}=c^{ew}(q \bar q,~ {\rm gauge})_H + c^{SQCD}(q\bar q) ~~
\label{cqq-gluino}
\eq
  in (\ref{universal-amp-gluino-cor}) where  $H=L ~{\rm or} ~R$, and
\bq
c^{\rm SQCD}(q\bar q)=
{1\over2}~c^{SQCD}(\tilde{q}\tilde{\bar q})=
-{\alpha_s\over3\pi}\Big [\ln{\s\over M^2_{SUSY}}\Big ]~~, \label{c-SUSY-QCD}
\eq
describes    the SUSY QCD correction, while
\bq
c^{\rm ew}(q\bar q, ~{\rm gauge})_H={\alpha\over4\pi} \Big [{I_q(I_q+1)\over s^2_W}
+{Y^2_q\over4c^2_W}\Big ]\Big (2 \ln{\s\over m^2_W}-\ln^2{\s\over m^2_W}\Big )~~,
\label{cew-qq-gauge}
\eq
gives the purely EW one. Here  $(I_q, ~Y_q)=(1/2, ~2/6)$ should be  used for
 $H=L $; and $(I_q, ~Y_q)=(0, ~2 Q_q)$ for  $H=R$, with $Q_q$ being
the quark charge. Since we  neglect quark
masses,  the associated Yukawa contributions are also neglected
in (\ref{universal-amp-gluino-cor}),
which is legitimate for the quarks found as partons
inside the proton.

Finally, the universal correction due to the final neutralino appearing in
(\ref{universal-amp-gluino-cor}),  is given by \cite{BRV, Denner, qqLHC}
\bq
c^{\rm ew}(W)= c^{\rm ew}(\tilde W)=
{\alpha \over 4 \pi  \swd }\Big [-\ln^2{\s \over M^2}\Big ] \label{cew-neutralino}
~~,
\eq
which is solely  induced  by the Wino component of the neutralino. \\

The only other EW correction that appear within the 1-loop LL level, is the
angular one,  given by \cite{BRV, Denner, qqLHC}
\bqa
F^{\rm ang}_{-++-}&=&  -(2I^3_q){eg_s\alpha\over2\pi s^3_W}Z^N_{2i}
\Big [\ln\Big ( \frac{\s}{\mwd}\Big )\Big ]
\Big [ \ln{-\t\over \s}+\ln{-\u\over \s}\Big ]
\Big ({\lambda^l\over2}\Big )_{ c_2  c_1} ~~, \nonumber \\
F^{\rm ang}_{-+-+}&=& (2I^3_q){eg_s\alpha\over2\pi s^3_W}Z^{N*}_{2i}
\Big [\ln \Big ( \frac{\s}{\mwd}\Big ) \Big ]
\Big [\ln{-\t\over \s}+ \ln{-\u\over \s}\Big ]
\Big ({\lambda^l\over2}\Big )_{ c_2  c_1} ~~,
\label{angular-amp-gluino-cor}
\eqa
which  arises solely
from gauge exchanges between the neutralino line,
(of which only the Wino component contributes),
and either the quark or the antiquark lines \cite{BRV, Denner, qqLHC}.

No one loop EW Renormalization Group (RG)  corrections are generated
in this case \cite{BRV, Denner, qqLHC}.

By adding to the Born helicity amplitudes in
(\ref{Born-amp-gluino-app}),
the corrections (\ref{universal-amp-gluino-cor})
and (\ref{angular-amp-gluino-cor}),   the complete
helicity amplitudes are constructed,  including  all  LL 1-loop
electroweak effects.

Taking into account all above corrections at asymptotic energies,
the dominant     amplitudes may be written as
\bqa
F^{\rm asym}_{-++-} & \simeq & \Big ({\lambda^l\over2}\Big )_{ c_2  c_1} 2 e g_s
\Bigg \{
\Big [ \frac{Z^N_{1i} }{6 \cw} +\frac{I^{(3)}_q}{\sw} Z^N_{2i} \Big ]
[1+\R_{\tilde g L}]+
\frac{I^{(3)}_q}{\sw}  Z^{N}_{2i}c^{ew}(\chi) \nonumber \\
&& -I^{(3)}_q{\alpha\over2\pi s^3_W}Z^N_{2i}
\Big [\ln\Big ( \frac{s}{\mwd}\Big )\Big ]
\Big [ \ln{-t\over s}+\ln{-u\over s}\Big ] \Bigg \}~~, \nonumber \\
F^{\rm asym}_{-+-+} & \simeq & - \Big ({\lambda^l\over2}\Big )_{ c_2  c_1} 2 e g_s
\Bigg \{
\Big [ \frac{Z^{N*}_{1i} }{6 \cw} +\frac{I^{(3)}_q}{\sw} Z^{N*}_{2i} \Big ]
[1+\R_{\tilde g L}]+
\frac{I^{(3)}_q}{\sw}  Z^{N*}_{2i}c^{ew}(\chi) \nonumber \\
&& -I^{(3)}_q{\alpha\over2\pi s^3_W}Z^{N*}_{2i}
\Big [\ln\Big ( \frac{s}{\mwd}\Big )\Big ]
\Big [ \ln{-t\over s}+\ln{-u\over s}\Big ] \Bigg \}~~, \nonumber \\
F^{\rm asym}_{+--+} & \simeq &  \Big ({\lambda^l\over2}\Big )_{ c_2  c_1} 2 e g_s
 \frac{Q_q}{\cw} Z^{N*}_{1i} [1+\R_{\tilde g R}] ~~, \nonumber \\
 F^{\rm asym}_{+-+-} & \simeq & - \Big ({\lambda^l\over2}\Big )_{ c_2  c_1} 2 e g_s
 \frac{Q_q}{\cw} Z^{N}_{1i} [1+\R_{\tilde g R}] ~~. \label{qq-gluino-amp-asym}
\eqa\\

\section{Helicity amplitudes for $ q g \to \tchi^0_i \tilde q_{L,R} $ \label{squark-appendix}}

Defining momenta and helicities as in  (\ref{prosses-squark}),
the Born amplitudes in (\ref{Born-amp-squark}) lead to the helicity amplitudes
\bq
F^B_{\lambda_1 \mu_2;\tau_1}(q_{c1}g_l\to \tchi^0_i \tilde q_{Ic_2})
=F^{\rm B(a)}_{\lambda_1 \mu_2;\tau_1}(q_{c1}g_l\to \tchi^0_i \tilde q_{Ic_2})
+F^{\rm B(b)}_{\lambda_1 \mu_2;\tau_1}(q_{c1}g_l\to \tchi^0_i \tilde q_{Ic_2})
~~, \label{Born-amp-squark1-app}
\eq
with the separate contributions from the two diagrams in Figs.\ref{qg-squark-fig}a,b giving
\bqa
&& F^{\rm B(a)}_{\lambda_1 \mu_2;\tau_1}
=-\frac{g_s A^{0I}_i(\tilde q_I)}{2 s^{1/4}\sqrt{E_1+\mi}}
\Big (\frac{\lambda^l}{2}\Big )_{c_2c_1}
\Big \{(E_1+\mi +p)[\mu_2(\lambda_1-\tau_1)+ (\lambda_1-\tau_1)^2] \cos\frac{\theta}{2}
\nonumber \\
&&  + (E_1+\mi -p)[(\lambda_1+\tau_1)
+ \mu_2 (\lambda_1+\tau_1)^2] \sin\frac{\theta}{2} \Big \} \delta_{I\lambda_1} ~~, \nonumber \\
&& F^{\rm B(b)}_{\lambda_1 \mu_2;\tau_1} =\frac{g_s A^{0I}_i(\tilde q_I)} {2 \sqrt{(E_1+\mi)}}
~ \frac{s^{3/4}\beta \mu_2\sin\theta}{(t-m_{\tilde q_I}^2)}
\Big (\frac{\lambda^l}{2}\Big )_{c_2c_1}
\Big \{(E_1+\mi -p)(\lambda_1+\tau_1)^2 \cos\frac{\theta}{2} \nonumber \\
&&  - (E_1+\mi +p)(\lambda_1-\tau_1) \sin\frac{\theta}{2} \Big \} \delta_{I\lambda_1}
~~, \label{Born-amp-squark2-app}
\eqa
where by a slight abuse of notation, $\delta_{I\lambda_1}$ simply indicates that
$(I=L,~R) $,  is uniquely  associated with the quark helicity
being $( \lambda_1=-1/2, ~+1/2)$ respectively. An alternative expression might be obtained by
substituting in  (\ref{Born-amp-squark2-app})
\bq
\frac{E_1+m_i \pm p}{\sqrt{E_1+m_i}}=\frac{1}{\sqrt{2} \s^{1/4}}
\Big [ \sqrt{(\sqrt{\s}+m_i)^2-m_j^2} \pm
\sqrt{(\sqrt{\s}-m_i)^2-m_j^2} \Big ]~. \label{Born-amp-squark2-substitute}
\eq

At asymptotic energies (much higher than all masses), (\ref{Born-amp-squark1-app},
\ref{Born-amp-squark2-app}) imply that   there is  only
one  non-vanishing amplitude for each squark type; \ie
\bqa
I=L   ~~  \Rightarrow ~~
F^{\rm B~asym}_{-++}(q_{c1}g_l\to \tchi^0_i \tilde q_{Lc_2})
& \simeq & -g_s\sqrt{2} A^{0L}_i(\tilde q_L)
\Big (\frac{\lambda^l}{2}\Big )_{c_2c_1} \cos
\Big (\frac{\theta}{2}\Big )
\nonumber \\
& = &2 e g_s \Big [ \frac{Z^N_{1i} }{6 \cw}
+\frac{I^{(3)}_q}{\sw} Z^N_{2i} \Big ]
\Big (\frac{\lambda^l}{2}\Big )_{c_2c_1} \cos
\Big (\frac{\theta}{2}\Big )
~, \nonumber \\
I=R  ~~  \Rightarrow ~~
F^{\rm B~asym}_{+--}(q_{c1}g_l\to \tchi^0_i \tilde q_{Rc_2})
& \simeq & -g_s\sqrt{2} A^{0R}_i(\tilde q_R)
\Big (\frac{\lambda^l}{2}\Big )_{c_2c_1} \cos
\Big (\frac{\theta}{2}\Big )
\nonumber \\
& =& - 2 e g_s \frac{Q_q}{\cw} Z^{N*}_{1i}
\Big (\frac{\lambda^l}{2}\Big )_{c_2c_1} \cos
\Big (\frac{\theta}{2}\Big )
~, \label{Born-amp-asym-squark2}
\eqa
with the neutralino squark couplings in the r.h.s. determined
by (\ref{chi0-Lsquarks}, \ref{chi0-Rsquarks}).

As in the case of  \ref{gluino-appendix},
the 1-loop Universal EW and SUSY QCD  LL  corrections,
only affect the dominant  amplitudes in (\ref{Born-amp-asym-squark2}), and
they are associated to the   quark, squark, or  Wino component of the final neutralino
line. Defining now
\bq
\R_{\tilde q I}  \equiv c^{ew}(q\bar q, ~\rm{gauge})_{I}
+\frac{1}{2}[ c^{SQCD}(q\bar q)+c^{SQCD}(\tilde q \bar{\tilde q})] ~~, \label{cqq-squark}
\eq
where (\ref{cew-qq-gauge},\ref{c-SUSY-QCD}) are used and the Yukawa
terms have again been neglected, the net resulting universal
correction  is
\bqa
F^{\rm Univ}_{I,-I;-I}(q g\to \tchi^0_i \tilde q_I)&=&
F^{\rm B~ asym}_{I,-I;-I}(q g\to \tchi^0_i \tilde q_I) \R_{\tilde q I}
\nonumber \\
& +& c^{ew}(\tilde W )(2I^3_q)Z^N_{2i}
\Big ({eg_s\over s_W}\Big )\Big ({\lambda^l\over2}\Big )_{c_2c_1}
\cos\Big ( {\theta\over2}\Big ) ~\delta_{I,L}~~,
\label{universal-amp-cor-squark}
\eqa
where (\ref{cew-qq-gauge}, \ref{cew-neutralino}, \ref{Born-amp-asym-squark2}) are used
for  $I=L~{\rm or}~R$.

The one loop  angular electroweak corrections are induced by
inserting  to the diagrams in Fig.\ref{qg-squark-fig}  either a $W$-exchange
between the neutralino and the $q$-leg, or a $(W,B)$ exchange between the
$q$ and $\tilde q$ legs. The first case induces a $\ln^2(-\t)$-term, while
the second a $\ln^2(-\u)$-one. The net result is
\bqa
F^{\rm ang }_{-++}(q_{c1}g_l\to \tchi^0_i \tilde q_{Lc_2})
 & \simeq &  -\frac{ e\alpha g_s}{4 \pi}
\Big (\frac{\lambda^l}{2}\Big )_{c_2c_1} \cos \Big (\frac{\theta}{2}\Big )
\Bigg \{  \frac{Z^N_{1i}(1+26 \cwd ) }{108 \cw^3 \swd}
[\ln^2(-\u) -\ln^2(\s)]
 \nonumber \\
 & + &\frac{Z^N_{2i} (2I^{(3)}_q)}{\sw^3} \Big [
  [\ln^2(-\t)-\ln^2(\s)]+
  \frac{(1-10  \cwd ) }{36 \cwd }[\ln^2(-\u) -\ln^2(\s)]  \Big ]\Bigg \} ~~,
\nonumber \\
F^{\rm ang }_{+--}(q_{c1}g_l\to \tchi^0_i \tilde q_{Rc_2})
& \simeq &  \frac{ e\alpha g_s Q_q^3}{2 \pi \cw^3}
\Big (\frac{\lambda^l}{2}\Big )_{c_2c_1} \cos \Big (\frac{\theta}{2}\Big )
Z^{N*}_{1i}[\ln^2(-\u) -\ln^2(\s)] ~~,
\label{angular-amp-cor-squark}
\eqa
and as in \ref{gluino-appendix}, there are  no  one loop EW RG
corrections.

By adding to the Born helicity amplitudes in
(\ref{Born-amp-squark1-app}),
the corrections (\ref{universal-amp-cor-squark})
and (\ref{angular-amp-cor-squark}),   the complete
helicity amplitudes are constructed,  including  all  LL 1-loop
electroweak effects. At asymptotic energies these acquire the form
\bqa
&& F^{\rm asym }_{-++}(q_{c1}g_l\to \tchi^0_i \tilde q_{Lc_2})
 \simeq     e g_s
\Big (\frac{\lambda^l}{2}\Big )_{c_2c_1} \cos \Big (\frac{\theta}{2}\Big )
\Bigg \{ \Big [\frac{Z^N_{1i}}{3 \cw}+\frac{2 I^{(3)}_q Z^N_{2i}}{\sw} \Big ]
[1+\R_{\tilde q L}]
\nonumber \\
&&   -\frac{\alpha}{4 \pi \swd}\Bigg [ \frac{2 I^{(3)}_q Z^N_{2i}}{\sw}
\Big ( \ln^2(s)+ \ln^2(-t)-\ln^2(s)+
\frac{(1-10 \cwd)}{36 \cwd}[\ln^2(-u)-\ln^2(s)] \Big )
\nonumber \\
 &&  + \frac{Z^N_{1i}(1+26 \cwd)}{108 \cw^3} [\ln^2(-u)-\ln^2(s)] \Bigg ] \Bigg \}~,
\nonumber \\
&& F^{\rm asym }_{+--}(q_{c1}g_l\to \tchi^0_i \tilde q_{Rc_2})
 \simeq     -2 e g_s \frac{Q_q}{\cw} Z^{N*}_{1i}
\Big (\frac{\lambda^l}{2}\Big )_{c_2c_1} \cos \Big (\frac{\theta}{2}\Big )
\Big \{ 1+\R_{\tilde q R}
\nonumber \\
&& - \frac{\alpha Q_q^2 }{4\pi \cwd} [\ln^2(-u)-\ln^2(s)] \Big \}~~.
\label{amp-asym-squark}
\eqa\\

\section{ Helicity amplitudes for $ u \bar d \to \tchi^0_i \tchi^+_j$ \label{chargino-appendix} }

\begin{figure}[b]
\[
\hspace{-0.5cm}\epsfig{file=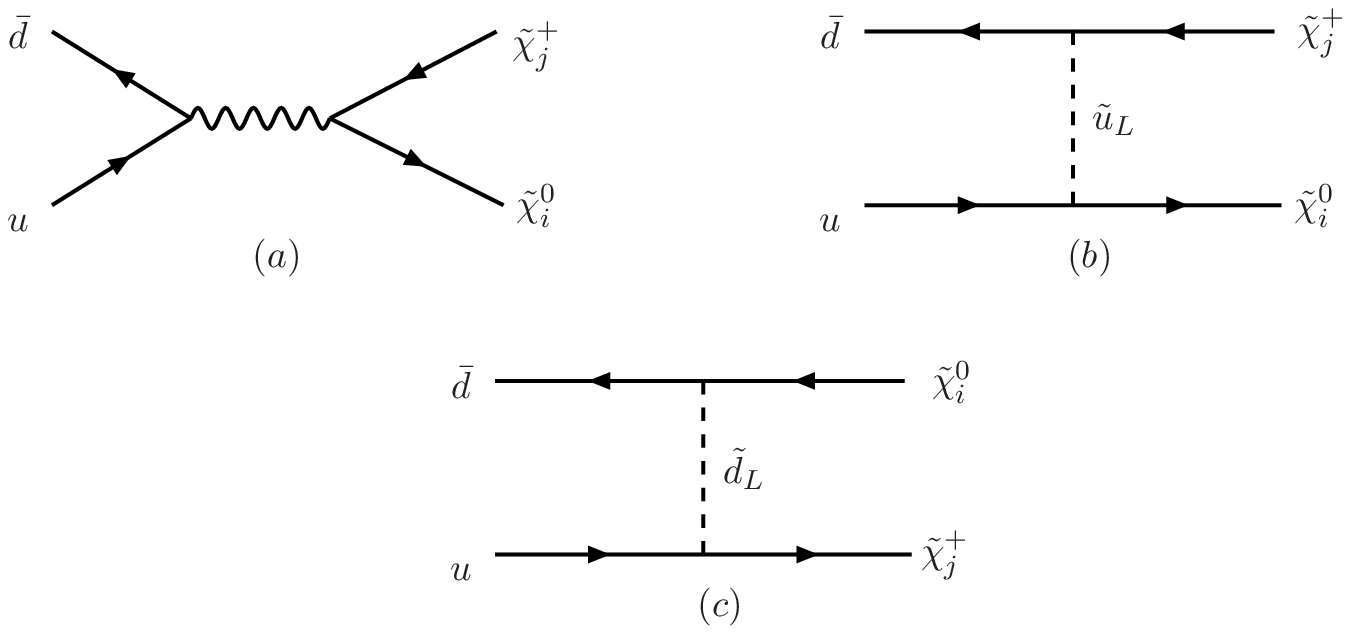,height=6.5cm, width=11cm}
\]
\caption[1]{Born diagrams for  $u \bar d  \to \tchi^0_i  \tchi^+_j $. }
\label{qq-chargino-fig}
\end{figure}

\begin{figure}[p]
\vspace*{-1cm}
\[
\hspace{-1.cm}\epsfig{file=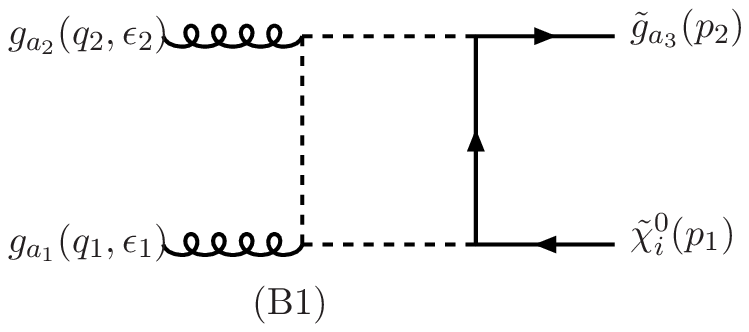,height=3.cm, width=6.5cm}
\hspace{2.cm}\epsfig{file=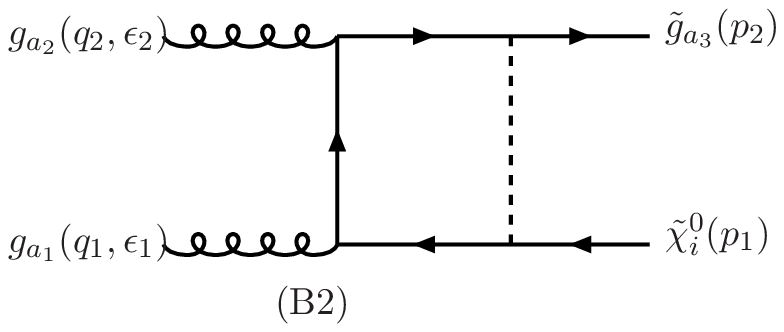,height=3.cm, width=6.5cm}
\]
\[
\hspace{0.5cm}\epsfig{file=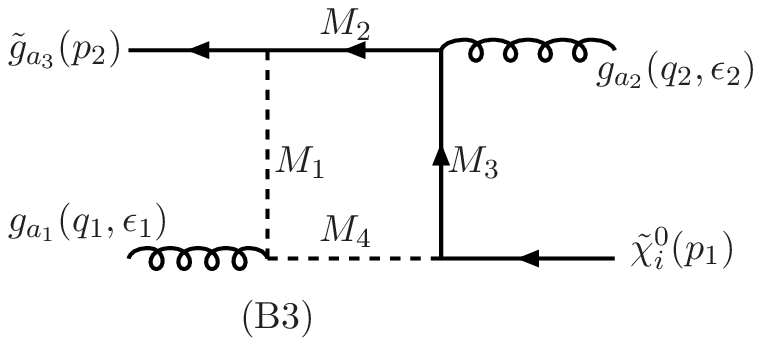,height=3.cm, width=6.5cm}
\]
\vspace*{1.cm}
\[
\hspace{-1.cm}\epsfig{file=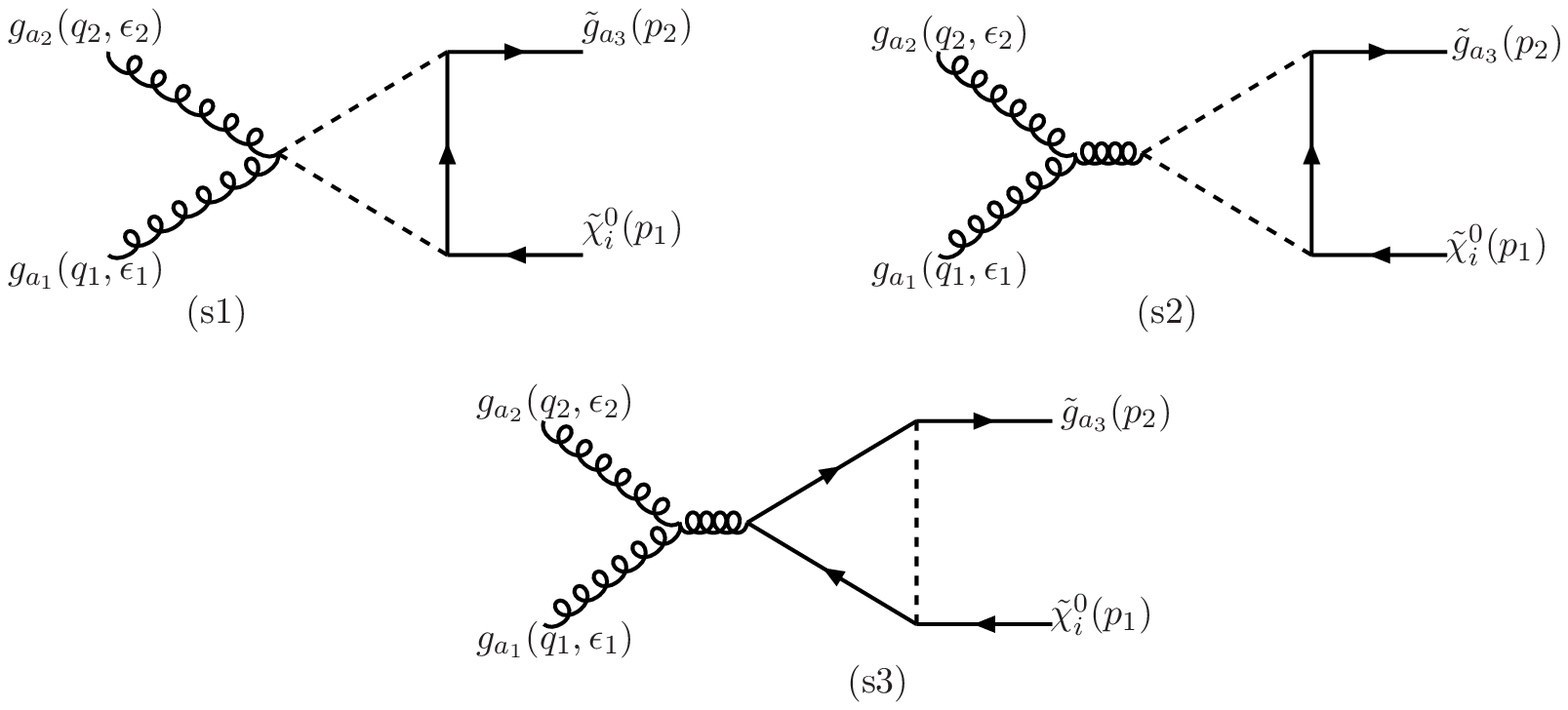,height=7.cm, width=13.cm}
\]
\vspace*{1.cm}
\[
\hspace{-1.cm}\epsfig{file=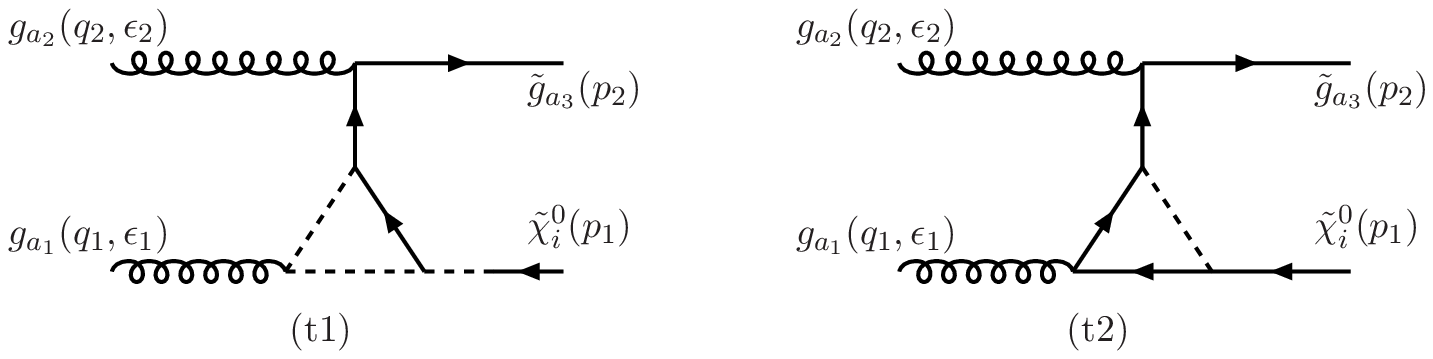,height=3.cm, width=13.cm}
\]
\caption[1]{Feynman diagrams for  $gg\to \tchi^0_i \tilde g$. The full, broken and
wavy lines  respectively denote fermions, scalars  and gluons.}
\label{gg-gluino-fig}
\end{figure}

For the process in (\ref{prosses-chi0p}), the respective contribution
to the helicity amplitudes in (\ref{Born-amp-chi0p}) from the three diagrams
in Figs.\ref{qq-chargino-fig}a,b,c  are
 \bqa
 S^{ijB}_{-+-+}&=&{e^2\sqrt{\s} \over 2\sqrt{2}s^2_W(\s -m^2_W)}
 (1+\cos\theta)\Big [- \sqrt{\s -(m_i-m_j)^2}(O^{WR}_{ji}+O^{WL}_{ji})
\nonumber \\
& - & \sqrt{\s -(m_i+m_j)^2}(O^{WR}_{ji}-O^{WL}_{ji})
\Big ] \nonumber ~~, \\
S^{ijB}_{-++-}&=&{e^2\sqrt{\s} \over 2\sqrt{2}s^2_W(\s -m^2_W)}
(1-\cos\theta)\Big[- \sqrt{\s -(m_i-m_j)^2}(O^{WR}_{ji}+O^{WL}_{ji})
\nonumber \\
& + & \sqrt{\s -(m_i+m_j)^2}(O^{WR}_{ji}-O^{WL}_{ji})
 \Big ] \nonumber ~~, \\
S^{ijB}_{-+++}&=& {e^2 \over 2\sqrt{2}s^2_W(\s -m^2_W)}
\sin\theta \Big[-\sqrt{\s -(m_i-m_j)^2}(m_i+m_j)
(O^{WR}_{ji}+O^{WL}_{ji})
\nonumber \\
& - & \sqrt{\s -(m_i+m_j)^2}(m_i-m_j) (O^{WR}_{ji}-O^{WL}_{ji})\Big ] \nonumber ~~, \\
S^{ijB}_{-+--}&=& {e^2 \over 2\sqrt{2}s^2_W(\s -m^2_W)} \sin\theta
\Big[ -\sqrt{\s -(m_i-m_j)^2}(m_i+m_j)
(O^{WR}_{ji}+O^{WL}_{ji})
\nonumber \\
& + &\sqrt{\s -(m_i+m_j)^2}(m_i-m_j)(O^{WR}_{ji}-O^{WL}_{ji})\Big ]  ~~, \label{chi0p-Born-s}
\eqa
\bqa
T^{ijB}_{-+-+}&=& - \frac{A^{0L}_i(\tilde u_L) A^{cL*}_j(\tilde u_L)}{4(\t-m_{\tilde u_L}^2 )}
(1+\cos\theta)
\sqrt{\s} \Big [ \sqrt{ \s-(m_i-m_j)^2} -\sqrt{\s-(m_i+m_j)^2}\Big ]~, \nonumber \\
T^{ijB}_{-++-}&=& - \frac{A^{0L}_i(\tilde u_L) A^{cL*}_j (\tilde u_L)}{4(\t-m_{\tilde u_L}^2 )}
(1-\cos\theta) \sqrt{\s} \Big [ \sqrt{ \s-(m_i-m_j)^2}
+\sqrt{\s-(m_i+m_j)^2}\Big ]~,
\nonumber \\
T^{ijB}_{-+++}&=& - \frac{A^{0L}_i(\tilde u_L)
A^{cL*}_j(\tilde u_L)}{4(\t-m_{\tilde u_L}^2 )}
\sin\theta
\Big [(m_i+m_j)\sqrt{ \s-(m_i-m_j)^2} -(m_i- m_j)
\sqrt{ \s-(m_i+m_j)^2}\Big ]~,
\nonumber \\
T^{ijB}_{-+--}&=& - \frac{A^{0L}_i(\tilde u_L)
A^{cL*}_j(\tilde u_L)}{4(\t-m_{\tilde u_L}^2 )}
\sin\theta
\Big [(m_i+m_j)\sqrt{ \s-(m_i-m_j)^2}
\nonumber \\
& +& (m_i- m_j) \sqrt{ \s-(m_i+m_j)^2}\Big ]~,  \label{chi0p-Born-t}
\eqa
\bqa
U^{ijB}_{-+-+}&=&  \frac{A^{0L*}_i(\tilde d_L)
A^{cL}_j(\tilde d_L)}{4(\u-m_{\tilde d_L}^2  )}
(1+\cos\theta) \sqrt{\s} \Big [ \sqrt{ \s-(m_i-m_j)^2}
+\sqrt{\s-(m_i+m_j)^2}\Big ]~,
\nonumber \\
U^{ijB}_{-++-}&=&  \frac{A^{0L*}_i(\tilde d_L)A^{cL}_j
(\tilde d_L)}{4(\u-m_{\tilde d_L}^2  )}
(1-\cos\theta) \sqrt{\s} \Big [ \sqrt{ \s-(m_i-m_j)^2}
-\sqrt{ \s-(m_i+m_j)^2}\Big ]~,
\nonumber \\
U^{ijB}_{-+++}&=&  \frac{A^{0L*}_i(\tilde d_L)A^{cL}_j
(\tilde d_L)}{4(\u-m_{\tilde d_L}^2  )}
\sin\theta
\Big [(m_i+m_j)\sqrt{ \s-(m_i-m_j)^2} +(m_i- m_j)
\sqrt{ \s-(m_i+m_j)^2}\Big ]~,
\nonumber \\
U^{ijB}_{-+--}&=&  \frac{A^{0L*}_i(\tilde d_L)A^{cL}_j
(\tilde d_L)}{4(\u-m_{\tilde d_L}^2  )}
\sin\theta
\Big [(m_i+m_j)\sqrt{ \s-(m_i-m_j)^2}
\nonumber \\
& -&(m_i- m_j)\sqrt{ \s-(m_i+m_j)^2}\Big ]~.  \label{chi0p-Born-u}
\eqa

 At energies much higher than all masses,  only two non vanishing
helicity  amplitudes remain, which simplify to
\bqa
F^{ij B, ~\rm as}_{-+-+}&=& S^{ij B, ~\rm as}_{-+-+}
+U^{ij B, ~\rm as}_{-+-+}
~~,  \nonumber \\
F^{ij B, ~\rm as}_{-++-}&=& S^{ij B, ~\rm as}_{-++-}
+T^{ij B, ~\rm as}_{-++-}
~~, \label{chi0p-Born-asym1}
\eqa
with
\bqa
&& S^{ij B, ~\rm as}_{-+-+} \simeq  -\frac{e^2} {\sqrt{2}\swd}(1+\cos\theta)O^{WR}_{ji}
=  -\frac{e^2}{\sqrt{2}\swd}(1+\cos\theta)
[ Z^{N*}_{2i}Z^{-}_{1j}+{1\over\sqrt{2}}Z^{N*}_{3i}Z^{-}_{2j}]
~~, \nonumber \\
&& S^{ij B, ~\rm as}_{-++-} \simeq  -\frac{e^2} {\sqrt{2}\swd}(1-\cos\theta)O^{WL}_{ji}
= -\frac{e^2}{\sqrt{2}\swd}(1-\cos\theta)
[Z^{N}_{2i}Z^{+*}_{1j}-{1\over\sqrt{2}}Z^{N}_{4i}Z^{+*}_{2j} ] ~~, \nonumber \\
&& T^{ij B, ~\rm as}_{-++-}  \simeq   A^{0L}_i
(\tilde u_L)A^{cL*}_j(\tilde u_L)=
\frac{e^2}{3\sqrt{2} \swd \cw}(Z^N_{1i}\sw
+3 Z^N_{2i}\cw) Z^{+*}_{1j} ~~,
\nonumber \\
&& U^{ij B, ~\rm as}_{-+-+} \simeq
- A^{0L*}_i(\tilde d_L)A^{cL}_j(\tilde d_L)=
- \frac{e^2}{3\sqrt{2} \swd \cw}(Z^{N*}_{1i}\sw
-3 Z^{N*}_{2i}\cw) Z^{-}_{1j} ~~ . \label{chi0p-Born-asym2}
\eqa

As before, the Universal 1-loop, purely gauge, EW  LL corrections are
\bqa
 && F^{ij,~ \rm Univ}_{-+-+}=
F^{ijB ~ \rm as}_{-+-+}~
{\alpha(1+26c^2_W)\over144\pi s^2_Wc^2_W} \Big [ 2 \ln\Big (\frac{\s}{\mwd} \Big )
-\ln^2\Big (\frac{\s}{\mwd} \Big )  \Big ]
\nonumber\\
&& - {e^2 \over\sqrt{2}s^2_W} (1+\cos\theta)
\Bigg \{
- {\alpha \over2 \pi s^2_W}  Z^{N*}_{2i}Z^{-}_{1j}\ln^2 \Big (\frac{\s}{\mwd} \Big )
\nonumber \\
&& + \Big ({\alpha(1+2c^2_W)\over16\pi\sqrt{2} s^2_Wc^2_W}
\Big [2\ln \Big (\frac{\s}{\mwd} \Big )
-\ln^2\Big (\frac{\s}{\mwd} \Big ) \Big ]
-{3\alpha m^2_b \over8\pi\sqrt{2} s^2_W \mwd \cos^2\beta}
\ln \Big (\frac{\s}{\mwd} \Big )\Big )
Z^{N*}_{3i}Z^{-}_{2j}\Bigg \}\nonumber\\
&& + {\alpha e^2 \over12\pi\sqrt{2} s^4_Wc_W}
Z^{-}_{1j}[Z^{N*}_{1i}s_W-6Z^{N*}_{2i}c_W ] \ln^2\Big (\frac{\s}{\mwd} \Big )~~,
\label{Univ-chi0p-mpmp}\\
 && F^{ij,~ \rm Univ}_{-++-}=
F^{ijB ~ \rm as}_{-++-}~
{\alpha(1+26c^2_W)\over144\pi s^2_Wc^2_W} \Big [ 2 \ln\Big (\frac{\s}{\mwd} \Big )
-\ln^2\Big (\frac{\s}{\mwd} \Big )  \Big ]
\nonumber\\
&& - {e^2 \over\sqrt{2}s^2_W} (1 -\cos\theta)
\Bigg \{
- {\alpha \over2 \pi s^2_W}  Z^{N}_{2i}Z^{+*}_{1j}\ln^2 \Big (\frac{\s}{\mwd} \Big )
\nonumber \\
&& - \Big ({\alpha(1+2c^2_W)\over16\pi\sqrt{2} s^2_Wc^2_W}
\Big [2\ln \Big (\frac{\s}{\mwd} \Big )
-\ln^2\Big (\frac{\s}{\mwd} \Big ) \Big ]
-{3\alpha m^2_t \over8\pi\sqrt{2} s^2_W \mwd \sin^2\beta}
\ln \Big (\frac{\s}{\mwd} \Big )\Big )
Z^{N}_{4i}Z^{+*}_{2j}\Bigg \}\nonumber\\
&& - {\alpha e^2 \over12\pi\sqrt{2} s^4_Wc_W}
Z^{+*}_{1j}[Z^{N}_{1i}s_W +6Z^{N}_{2i}c_W ] \ln^2\Big (\frac{\s}{\mwd} \Big )~~.
\label{Univ-chi0p-mppm}
\eqa
The first term in both  expressions (\ref{Univ-chi0p-mpmp}, \ref{Univ-chi0p-mppm}) are
  due to the quark external lines, the second and third come from
 the s-channel diagram in Fig.\ref{qq-chargino-fig}a, while the last term
 in (\ref{Univ-chi0p-mpmp})  and (\ref{Univ-chi0p-mppm}) are induced from
 the u- and t-channel diagram in Figs.\ref{qq-chargino-fig}c and b, respectively.

Finally, the third term in both (\ref{Univ-chi0p-mpmp}) and (\ref{Univ-chi0p-mppm}),
is  a Yukawa contribution
induced by the higgsino components of the chargino and neutralino
produced through the diagram in Fig.\ref{qq-chargino-fig}a. This  Yukawa contribution
appears in (\ref{Univ-chi0p-mpmp},\ref{Univ-chi0p-mppm}),
in spite of the fact that the participating quarks are massless \cite{BRV}.

The SUSY QCD 1-loop universal LL corrections are
\bq
F^{ij,~ \rm SQCD}_{-+\mp \pm}=
F^{ijB ~ \rm as}_{-+\mp \pm} \Big [
 -\frac{\alpha_s}{3\pi} \ln\Big (\frac{\s}{M_S^2} \Big )   \Big ]~~,
 \label{SQCD-chi0p}
\eq
while, in this case, there exist also a 1-loop RG single-log contribution,
caused by the $W^+$-exchange
in Fig.\ref{qq-chargino-fig}a, which is \cite{BRV}
\bq
F^{ij,~ \rm RG}_{-+\mp \pm}=
F^{ijB ~ \rm as}_{-+\mp \pm} \Big [
 \frac{\alpha }{4  \pi \swd}  \ln\Big (\frac{\s}{M_S^2} \Big )   \Big ]~~.
 \label{RG-chi0p}
\eq

Finally  the 1-loop LL EW angular corrections  are
\bqa
F^{ang}_{-+-+}&=&
{e^2\alpha\over 8 \pi} (1+\cos\theta)
\Bigg \{ \frac{\sqrt{2}}{\sw^4 } O^{WR}_{ji}
\Big [ \ln^2 \Big | \frac{\t}{M^2} \Big |+\ln^2\Big |
\frac{\u}{M^2} \Big |
-2  \ln^2 \Big | \frac{\s}{M^2} \Big | \Big ] \nonumber \\
& - & \frac{Z^{-}_{2j} Z^{N*}_{3i}}{3 \swd \cwd}
 \Big [ \ln^2\Big | \frac{\t}{M^2} \Big |-\ln^2\Big |
\frac{\u}{M^2} \Big | \Big ]
 \Bigg \}  \nonumber\\
& + & {e^2\alpha\over 12 \sqrt{2}  \pi \sw^4 \cw }
\Bigg \{ Z^{-}_{1j} Z^{N*}_{1i} \sw
\Big [ \ln^2 \Big | \frac{\t}{M^2} \Big |+\ln^2\Big |
\frac{\u}{M^2} \Big |
-2  \ln^2 \Big | \frac{\s}{M^2} \Big | \Big ] \nonumber \\
&  - & 6 \cw Z^{-}_{1j} Z^{N*}_{2i}
\Big [ \ln^2\Big | \frac{\u}{M^2}\Big |-\ln^2 \Big |
\frac{\s}{M^2} \Big | \Big ]  \Bigg \}~~,
 \label{ang-chi0p-mpmp} \\
 F^{ang}_{-++-}&=&
{e^2\alpha\over 8 \pi} (1-\cos\theta)
\Bigg \{ \frac{\sqrt{2}}{\sw^4 } O^{WL}_{ji}
\Big [ \ln^2 \Big | \frac{\t}{M^2} \Big |+\ln^2\Big |
\frac{\u}{M^2} \Big |
-2  \ln^2 \Big | \frac{\s}{M^2} \Big | \Big ]
\nonumber \\
&  + & \frac{Z^{+*}_{2j} Z^{N}_{4i}}{3 \swd \cwd}
 \Big [ \ln^2\Big | \frac{\t}{M^2} \Big |-\ln^2\Big |
\frac{\u}{M^2} \Big | \Big ]
 \Bigg \}  \nonumber\\
& - & {e^2\alpha\over 12 \sqrt{2}  \pi \sw^4 \cw }
\Bigg \{ Z^{+*}_{1j}\Big [\sw  Z^{N}_{1i} +6 \cw  Z^{N}_{2i}\Big ]
\Big [ \ln^2 \Big | \frac{\t}{M^2} \Big | -\ln^2
\Big | \frac{\s}{M^2} \Big |\Big ]
 \nonumber \\
 & + &   \sw Z^{+*}_{1j} Z^{N}_{1i}
 \Big [ \ln^2\Big | \frac{\u}{M^2} \Big |-\ln^2
\Big | \frac{\s}{M^2} \Big |\Big ]
   \Bigg \}~~.  \label{ang-chi0p-mppm}
\eqa

The fully corrected helicity amplitudes are obtained by adding to the Born expressions
 (\ref{chi0p-Born-s}, \ref{chi0p-Born-t}, \ref{chi0p-Born-u}), the corrections
(\ref{Univ-chi0p-mpmp}, \ref{Univ-chi0p-mppm}), (\ref{SQCD-chi0p}),
(\ref{RG-chi0p}) and (\ref{ang-chi0p-mpmp}, \ref{ang-chi0p-mppm}).

\newpage

\newpage

\clearpage
\newpage

\begin{figure}[p]
\vspace*{-2cm}
\[
\hspace{-0.5cm}\epsfig{file=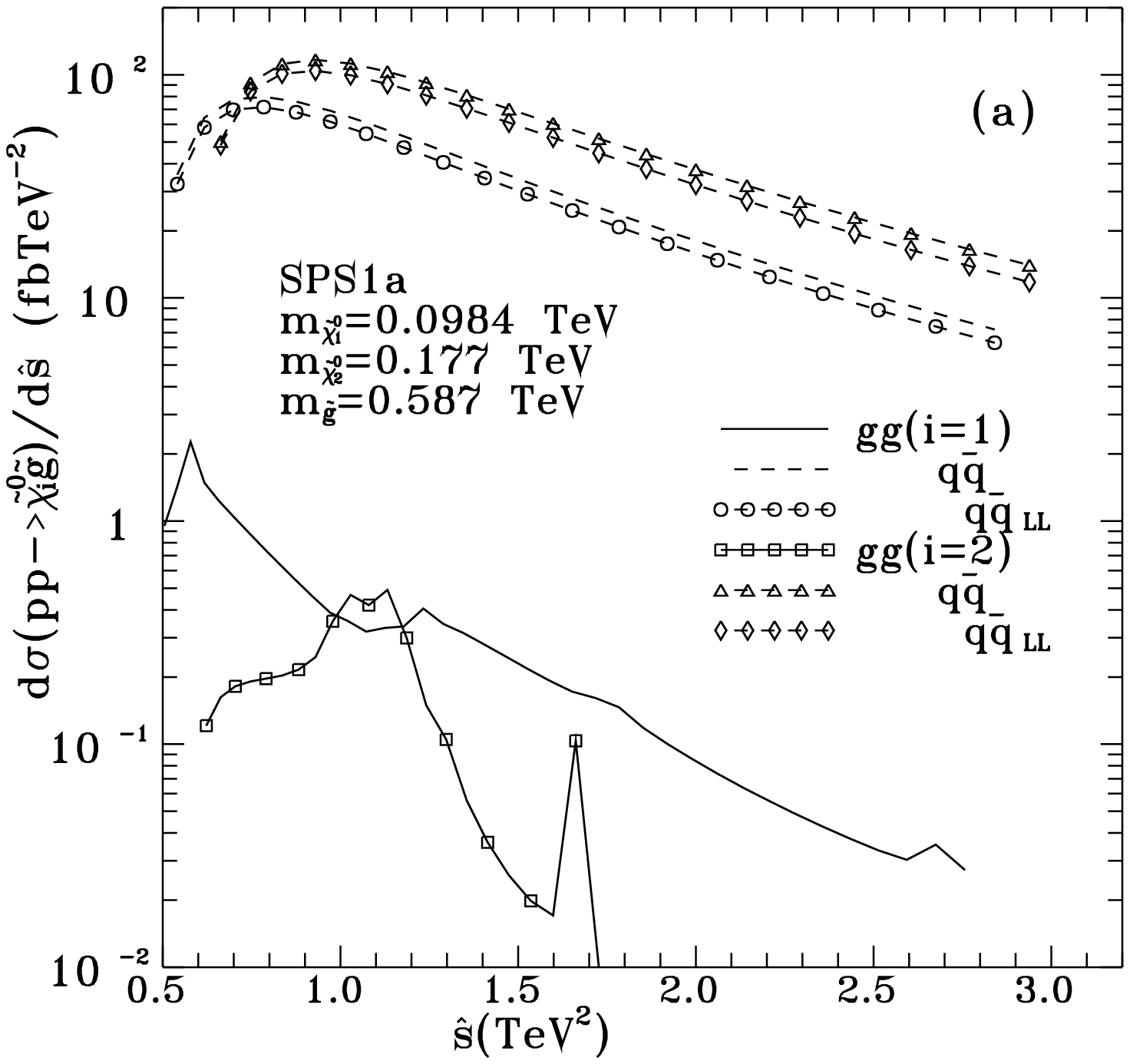,height=7.5cm, width=7.5cm}
\hspace{1.cm}\epsfig{file=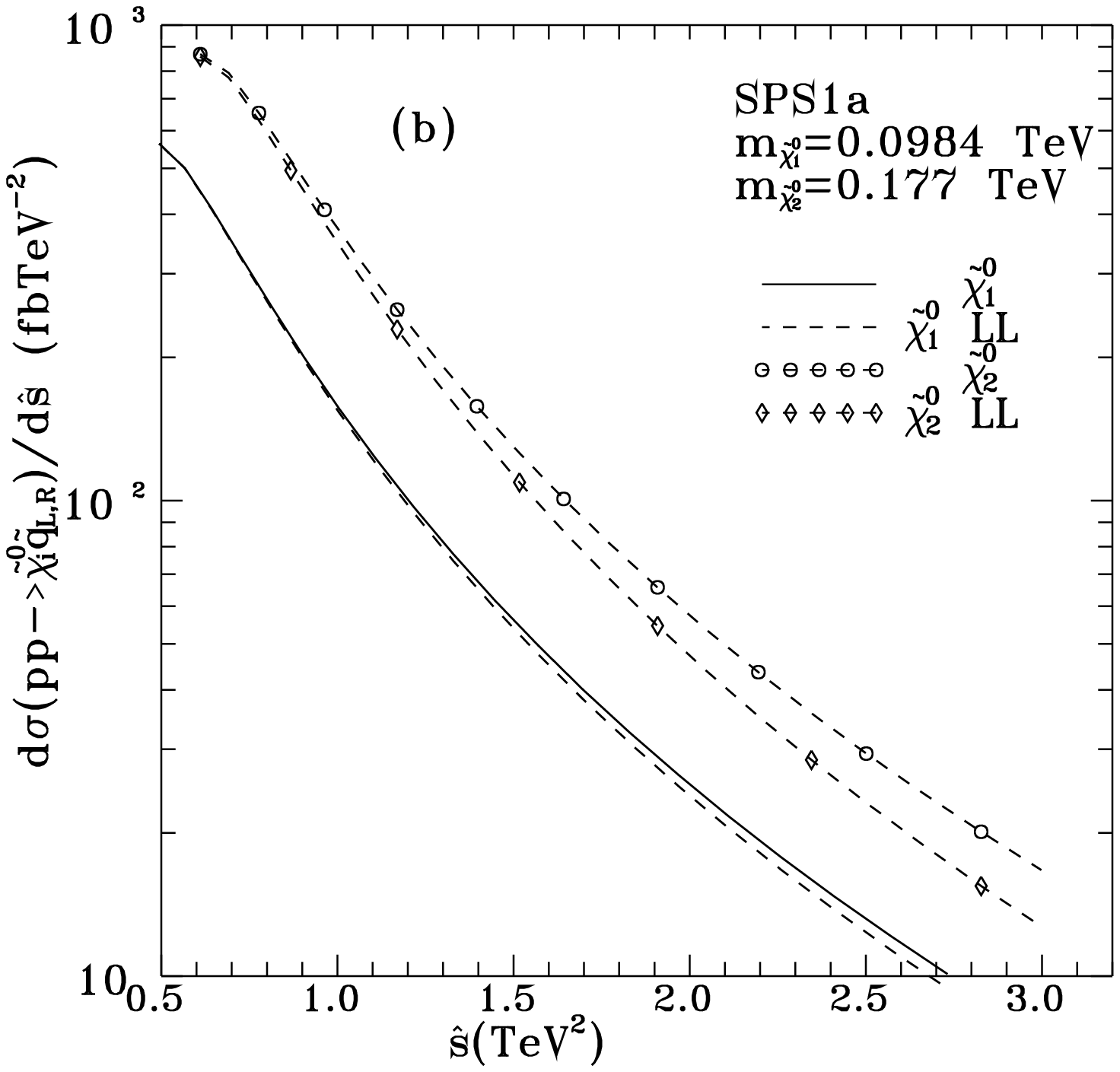,height=7.5cm, width=7.5cm}
\]
\[
\hspace{-0.5cm}\epsfig{file=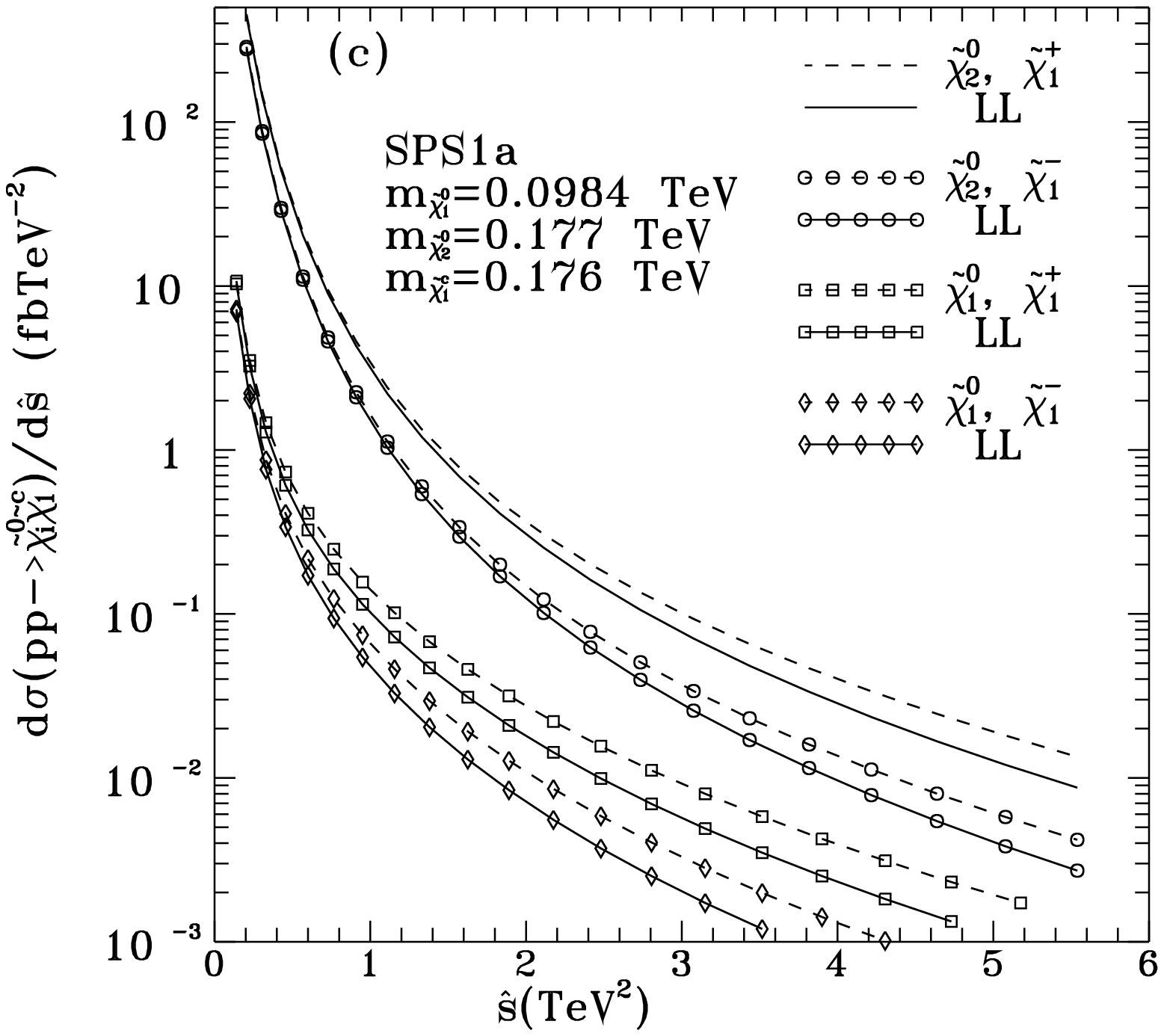,height=7.5cm, width=7.5cm}
\]
\caption[1]{SPS1a $\s$-distributions for $\tchi^0_{1,2}$ production, in association with
either $\tilde g$  or $\tilde q_{L,R}$ or $\tchi^\pm_1$; ($\tchi^c_j\equiv \tchi^\pm_j$).}
\label{SPS1a-mass-fig}
\end{figure}

\clearpage
\newpage

\begin{figure}[p]
\vspace*{-2cm}
\[
\hspace{-0.5cm}\epsfig{file=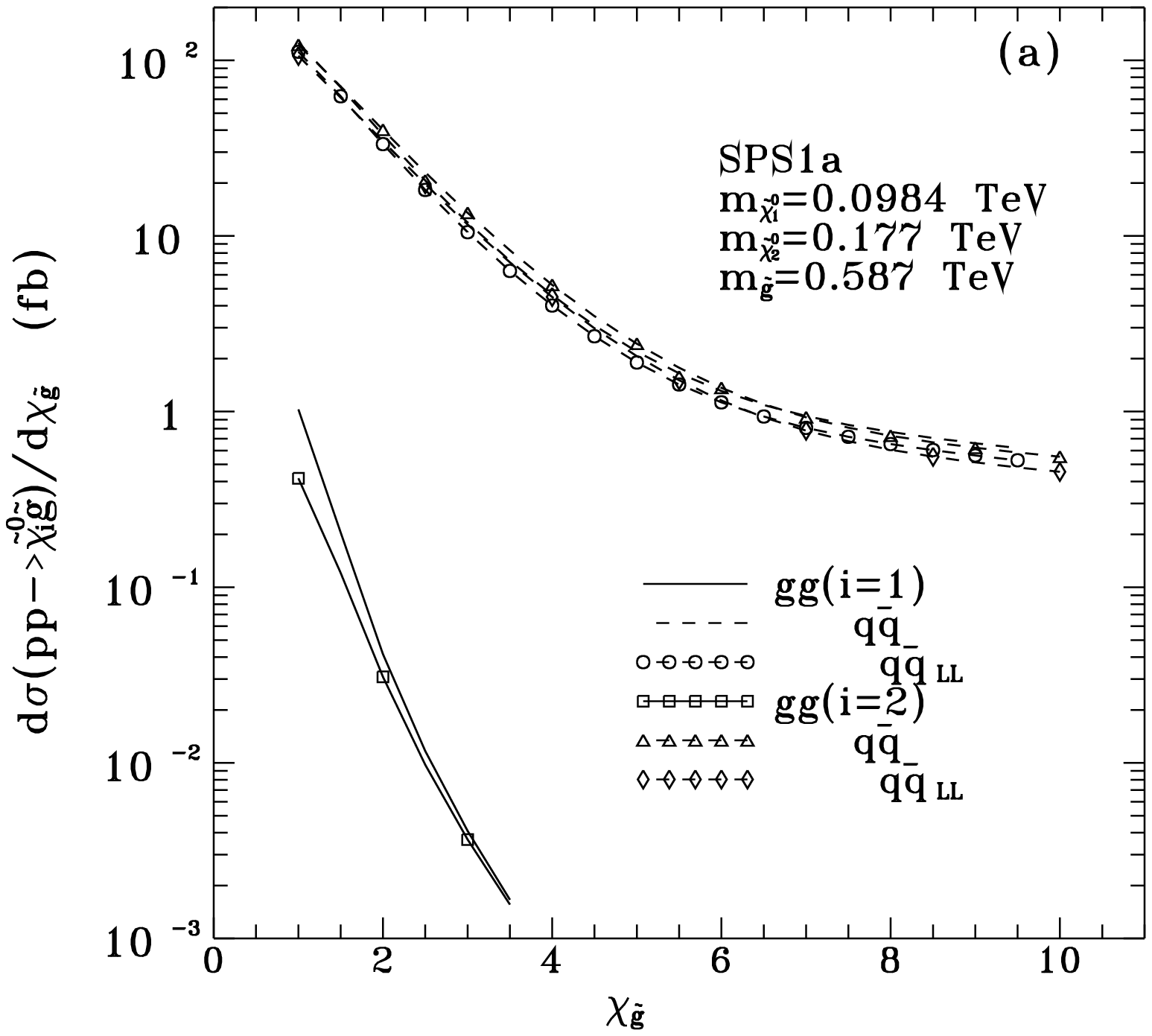,height=7.5cm, width=7.5cm}
\hspace{1.cm}\epsfig{file=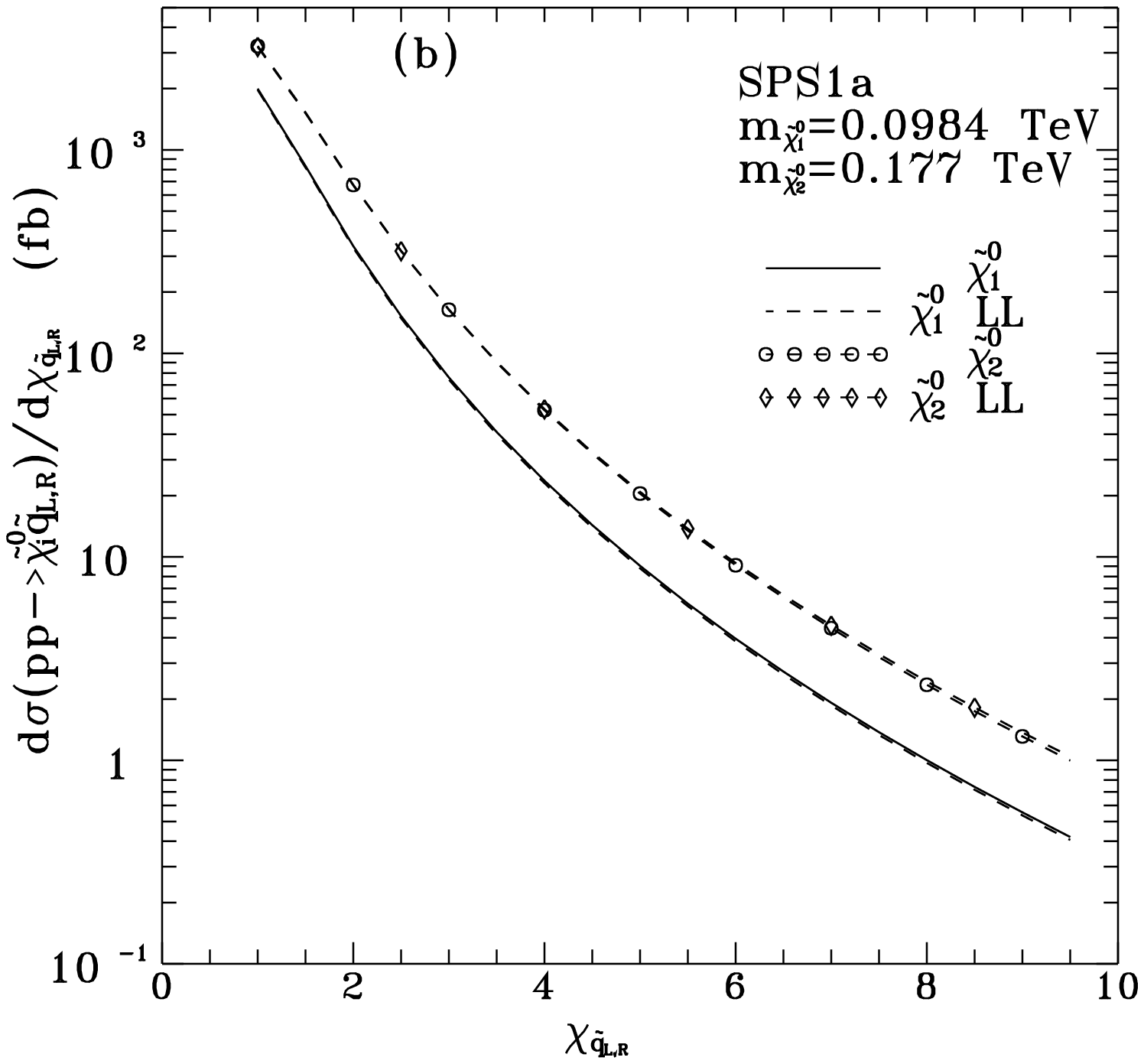,height=7.5cm, width=7.5cm}
\]
\[
\hspace{-0.5cm}\epsfig{file=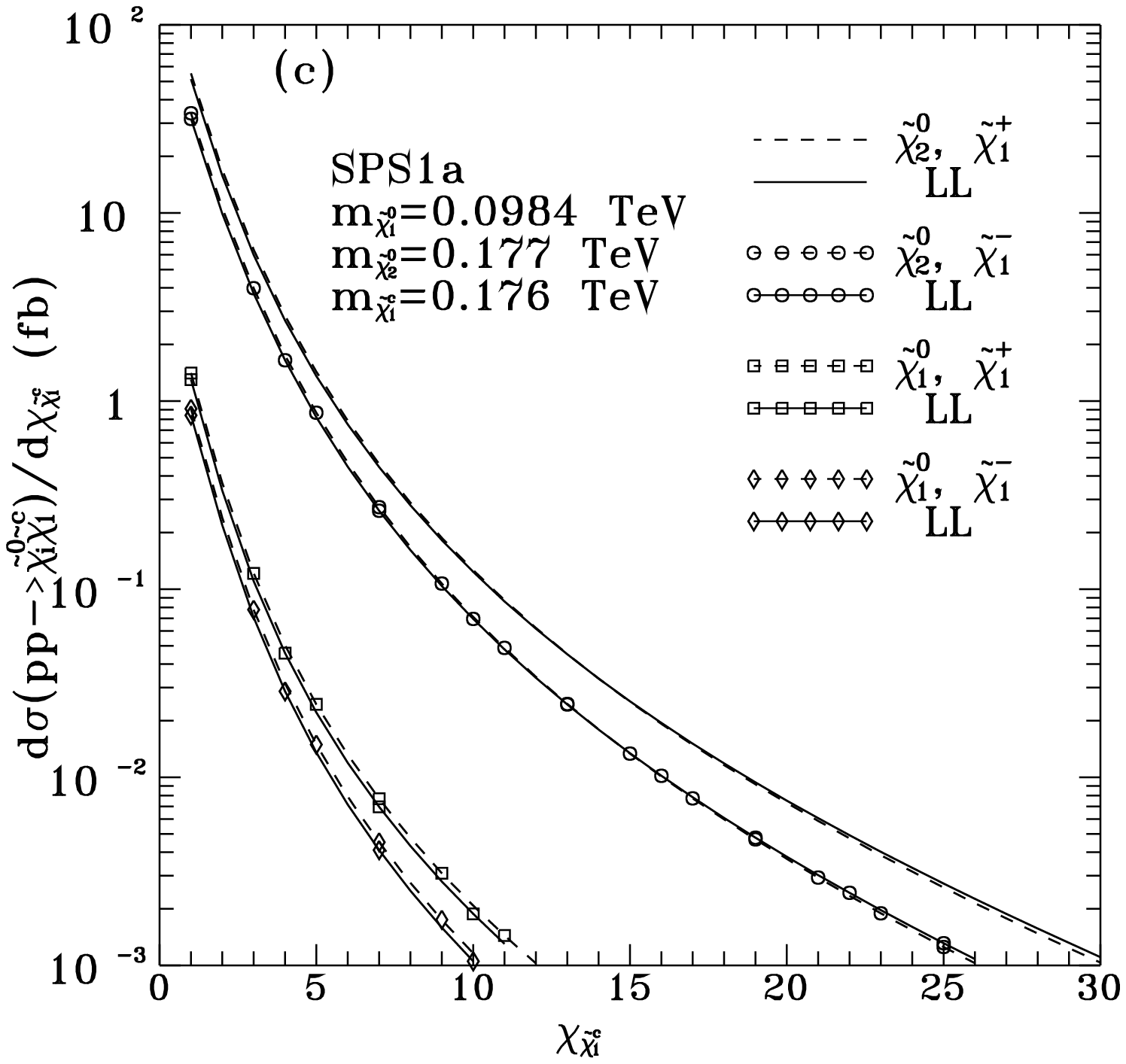,height=7.5cm, width=7.5cm}
\]
\caption[1]{SPS1a $\chi$-distributions for  $\tchi^0_{1,2}$ production, in association with
either $\tilde g$  or $\tilde q_{L,R}$ or $\tchi^\pm_1$; ($\tchi^c_j\equiv \tchi^\pm_j$).}
\label{SPS1a-chi-fig}
\end{figure}

\clearpage
\newpage

\begin{figure}[p]
\vspace*{-2cm}
\[
\hspace{-0.5cm}\epsfig{file=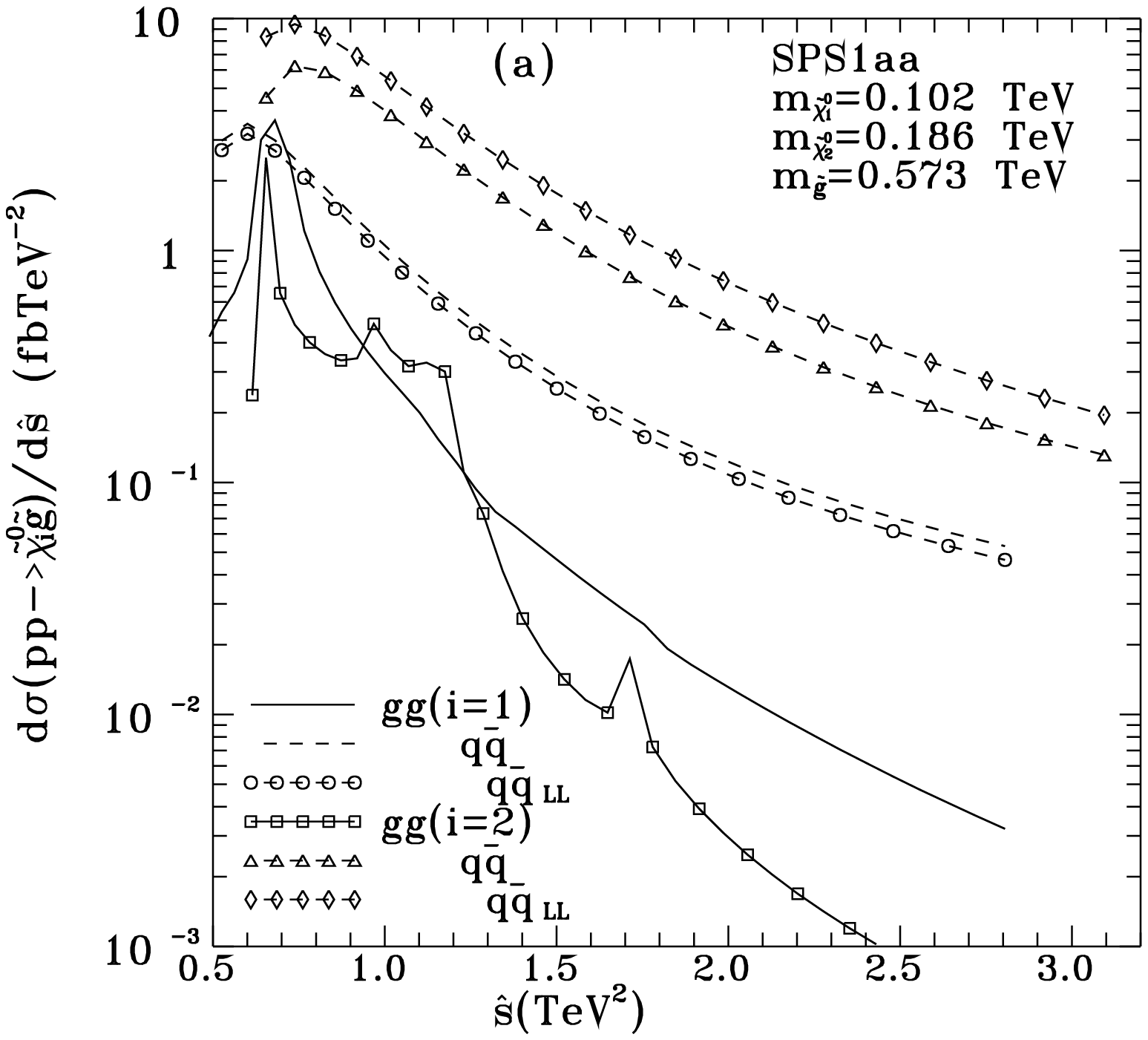,height=7.5cm, width=7.5cm}
\hspace{1.cm}\epsfig{file=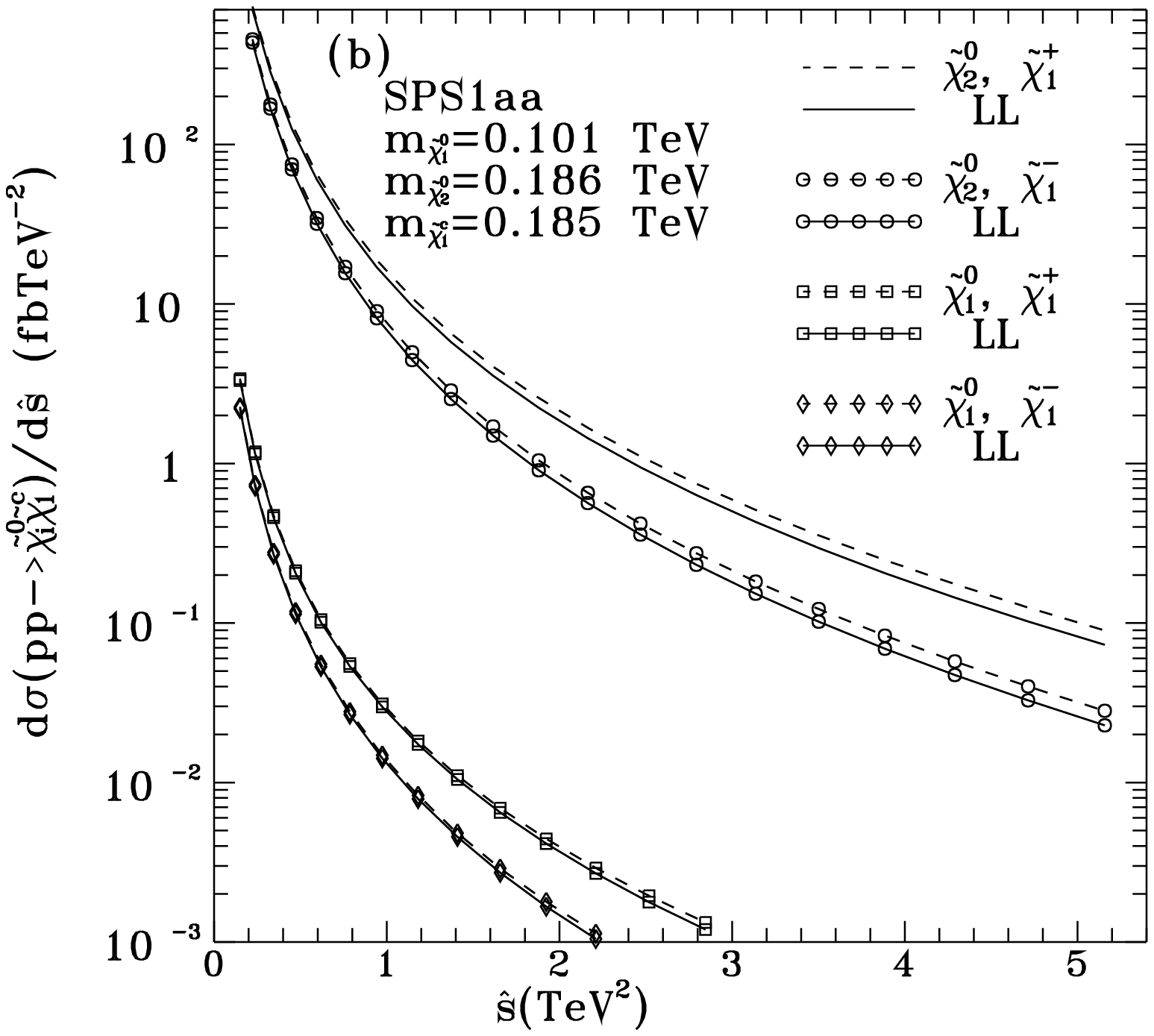,height=7.5cm, width=7.5cm}
\]
\[
\hspace{-0.5cm}\epsfig{file=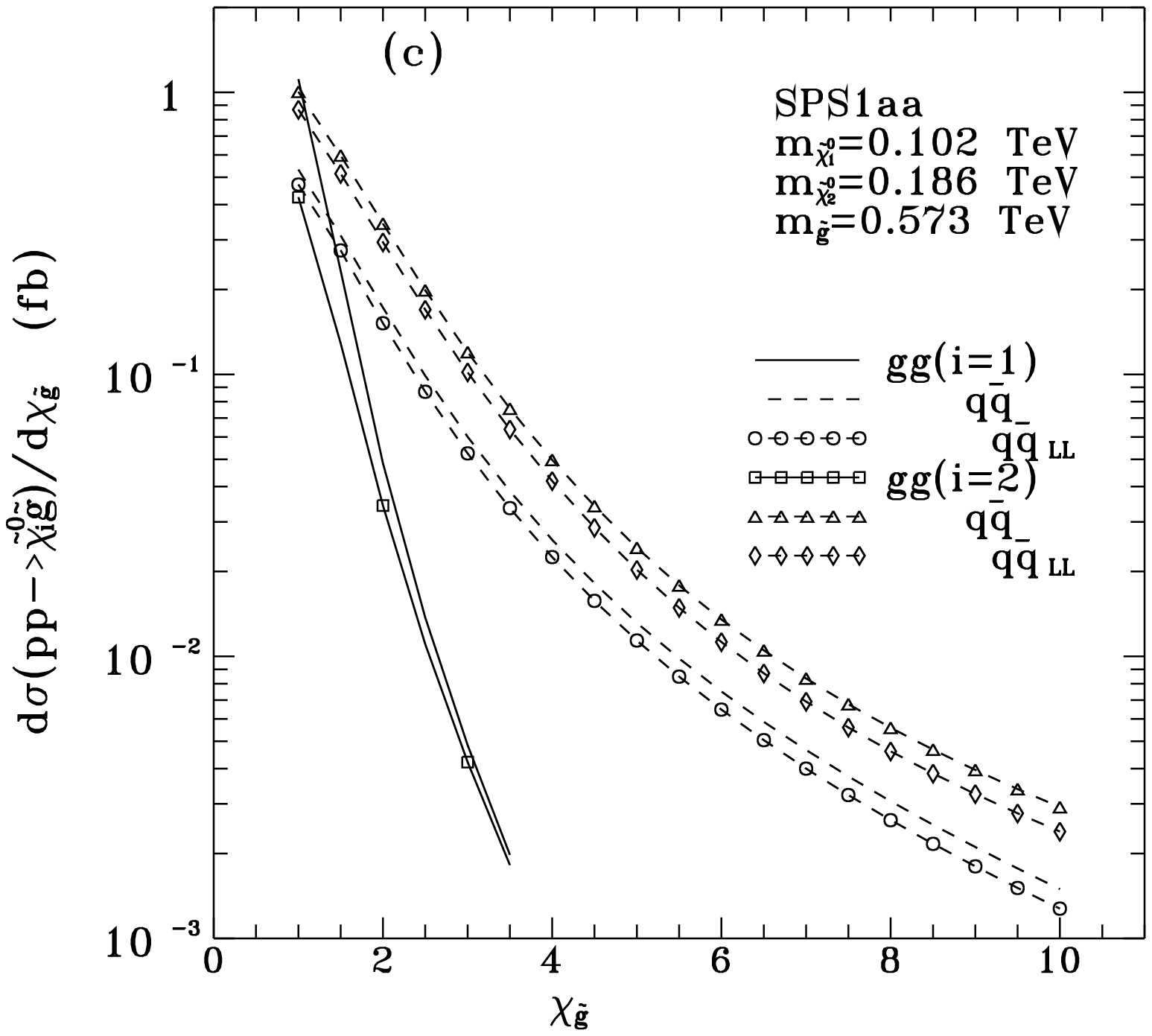,height=7.5cm, width=7.5cm}
\hspace{1.cm}\epsfig{file=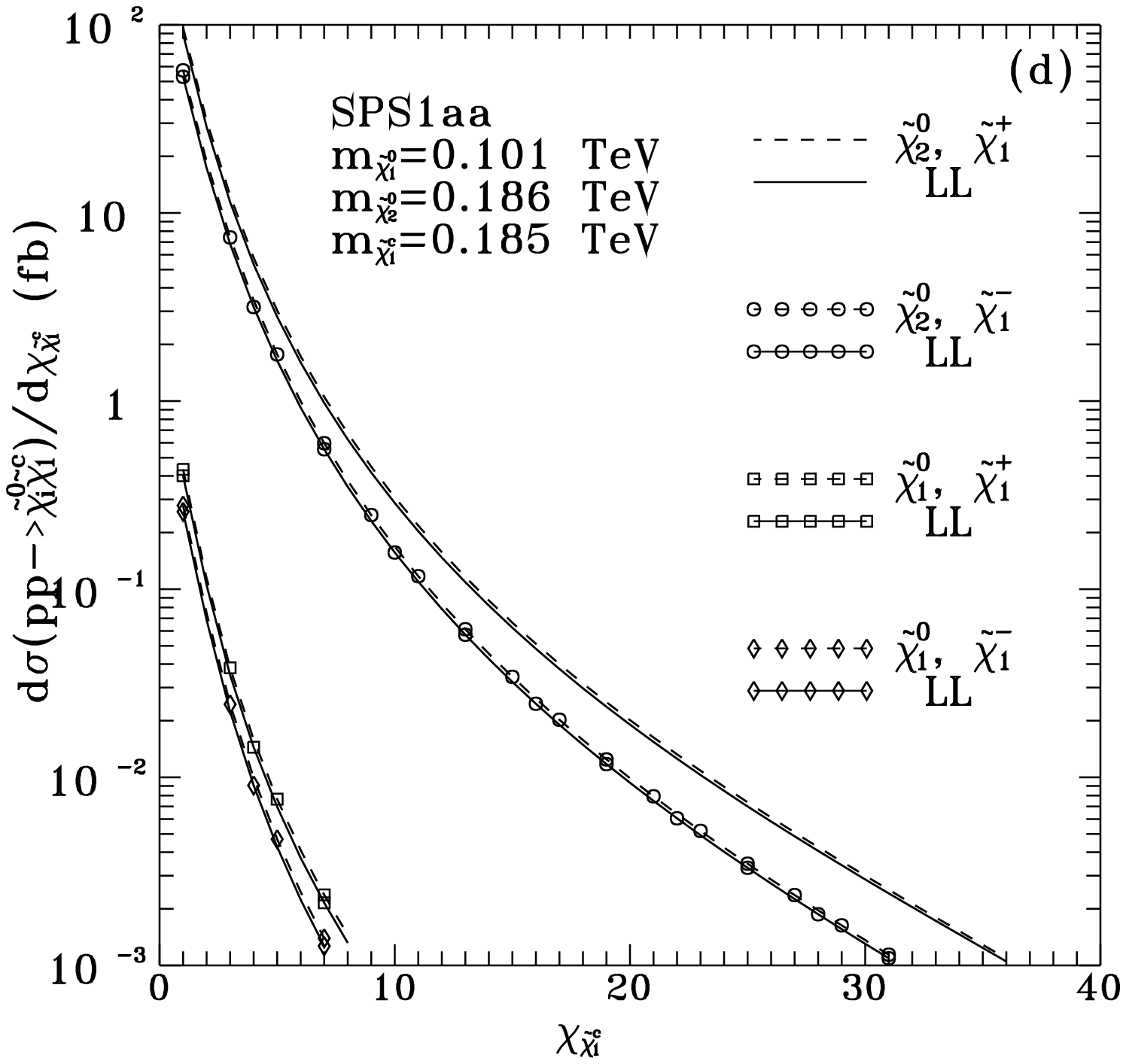,height=7.5cm, width=7.5cm}
\]
\caption[1]{SPS1aa $\s$- and $\chi$-distributions for $\tchi^0_{1,2}$ production
in association with either $\tilde g$  or $\tchi^\pm_1$.}
\label{SPS1aa-mass-chi-fig}
\end{figure}

\clearpage
\newpage

\begin{figure}[p]
\vspace*{-2cm}
\[
\hspace{-0.5cm}\epsfig{file=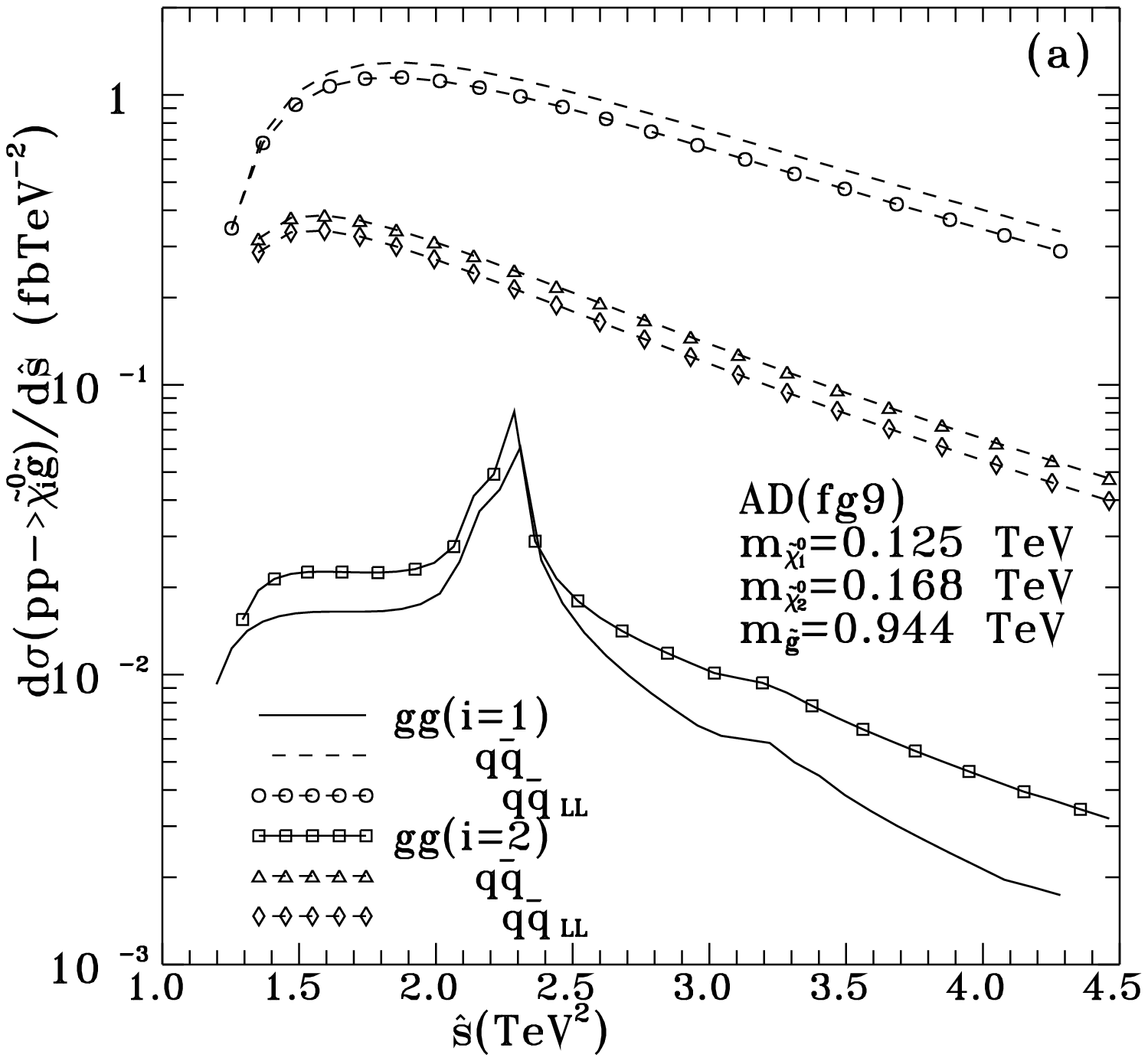,height=7.5cm, width=7.5cm}
\hspace{1.cm}\epsfig{file=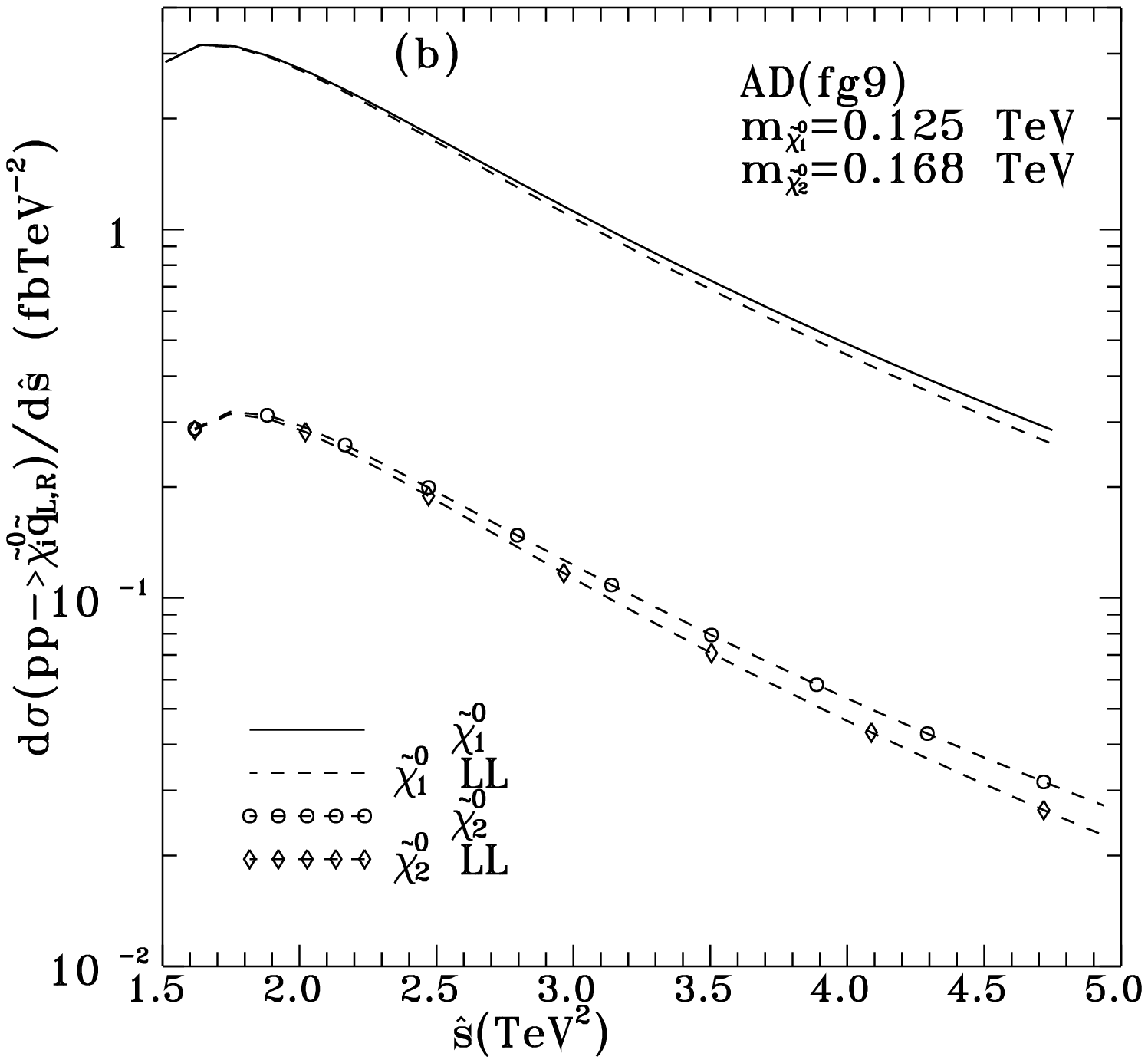,height=7.5cm, width=7.5cm}
\]
\[
\hspace{-0.5cm}\epsfig{file=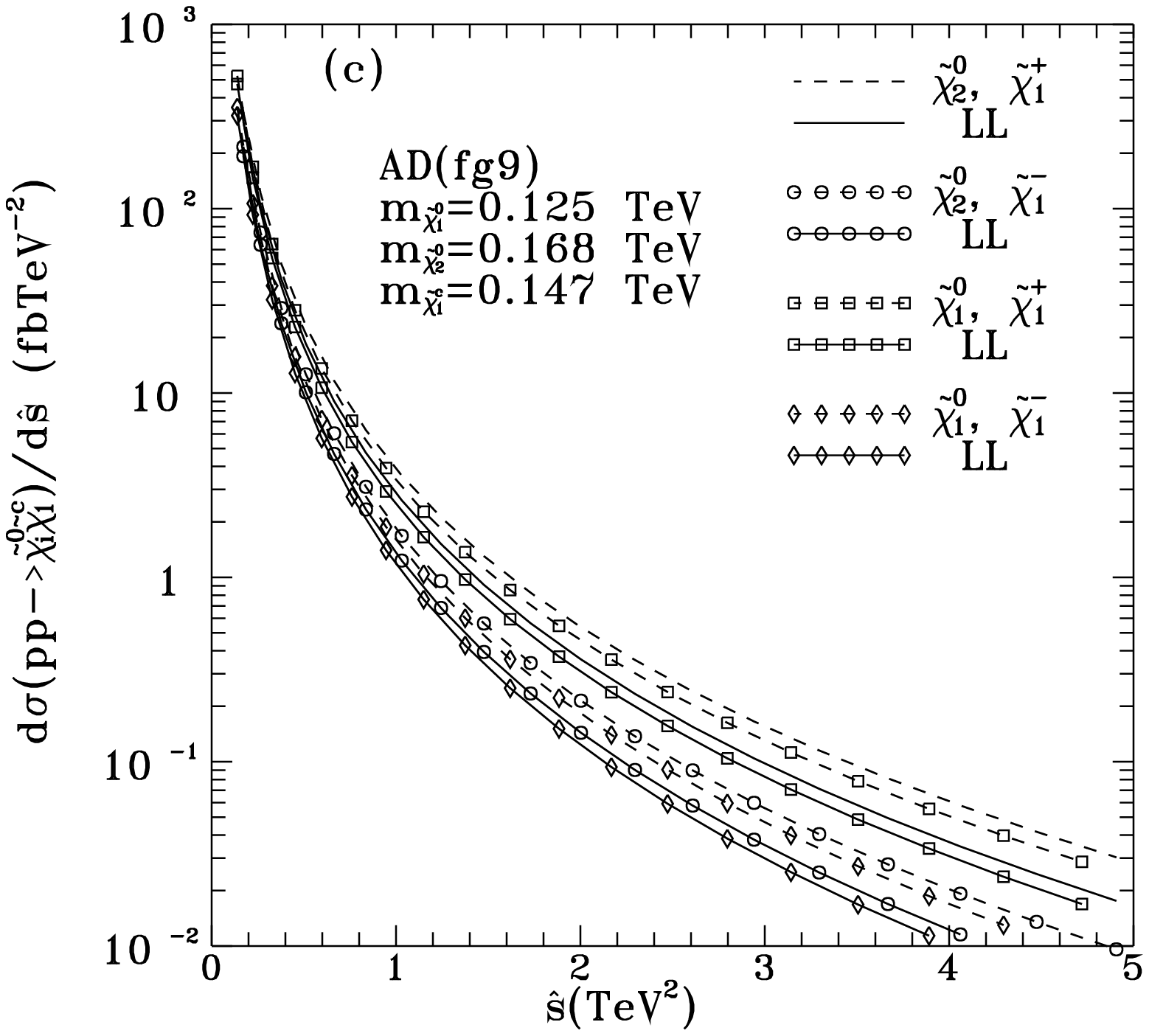,height=7.5cm, width=7.5cm}
\]
\caption[1]{AD(fg9)  $\s$-distributions for $\tchi^0_{1,2}$ production, in association with
either $\tilde g$  or $\tilde q_{L,R}$ or $\tchi^\pm_1$; ($\tchi^c_j\equiv \tchi^\pm_j$).}
\label{ADfg9-mass-fig}
\end{figure}

\clearpage
\newpage

\begin{figure}[p]
\vspace*{-2cm}
\[
\hspace{-0.5cm}\epsfig{file=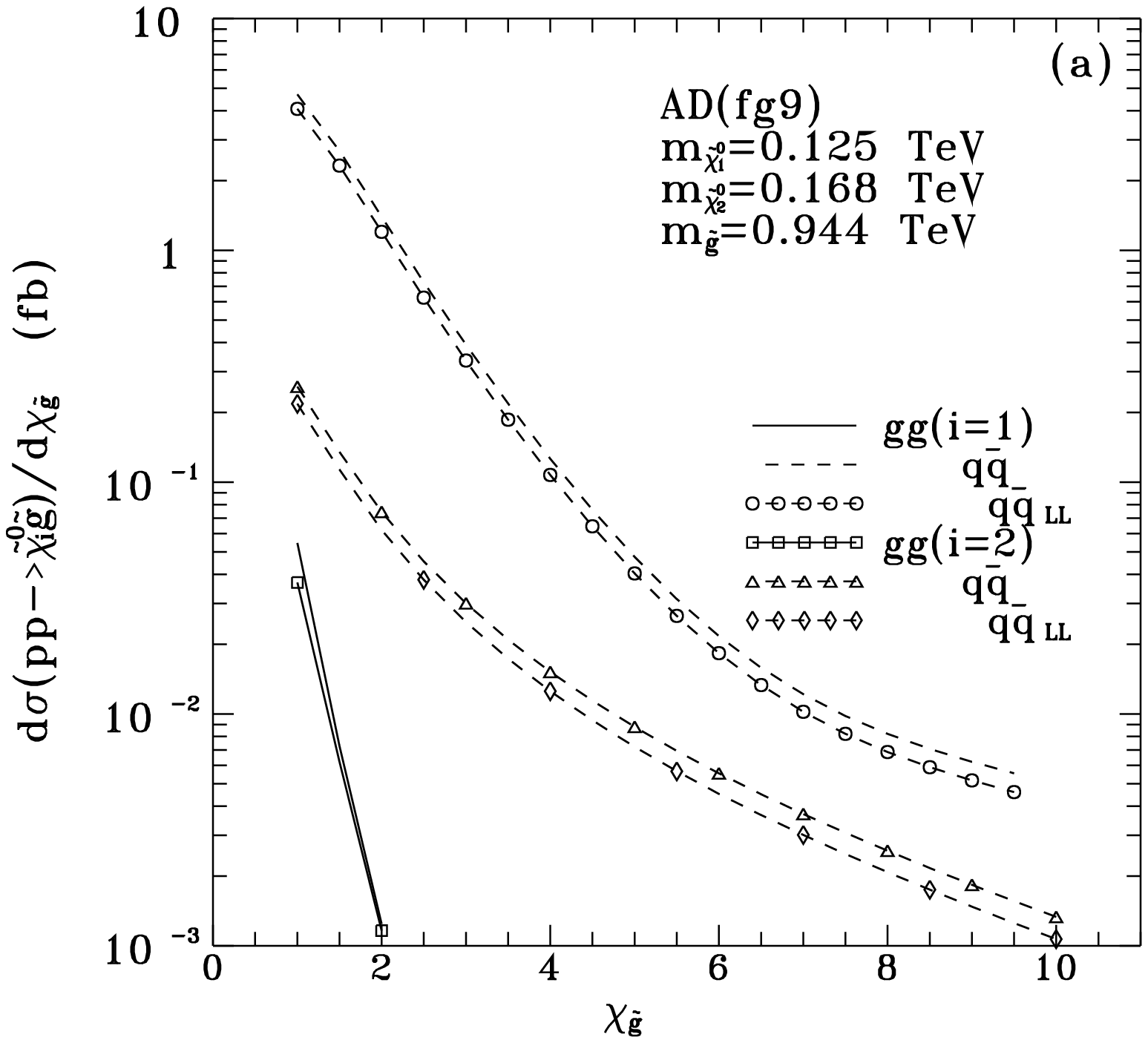,height=7.5cm, width=7.5cm}
\hspace{1.cm}\epsfig{file=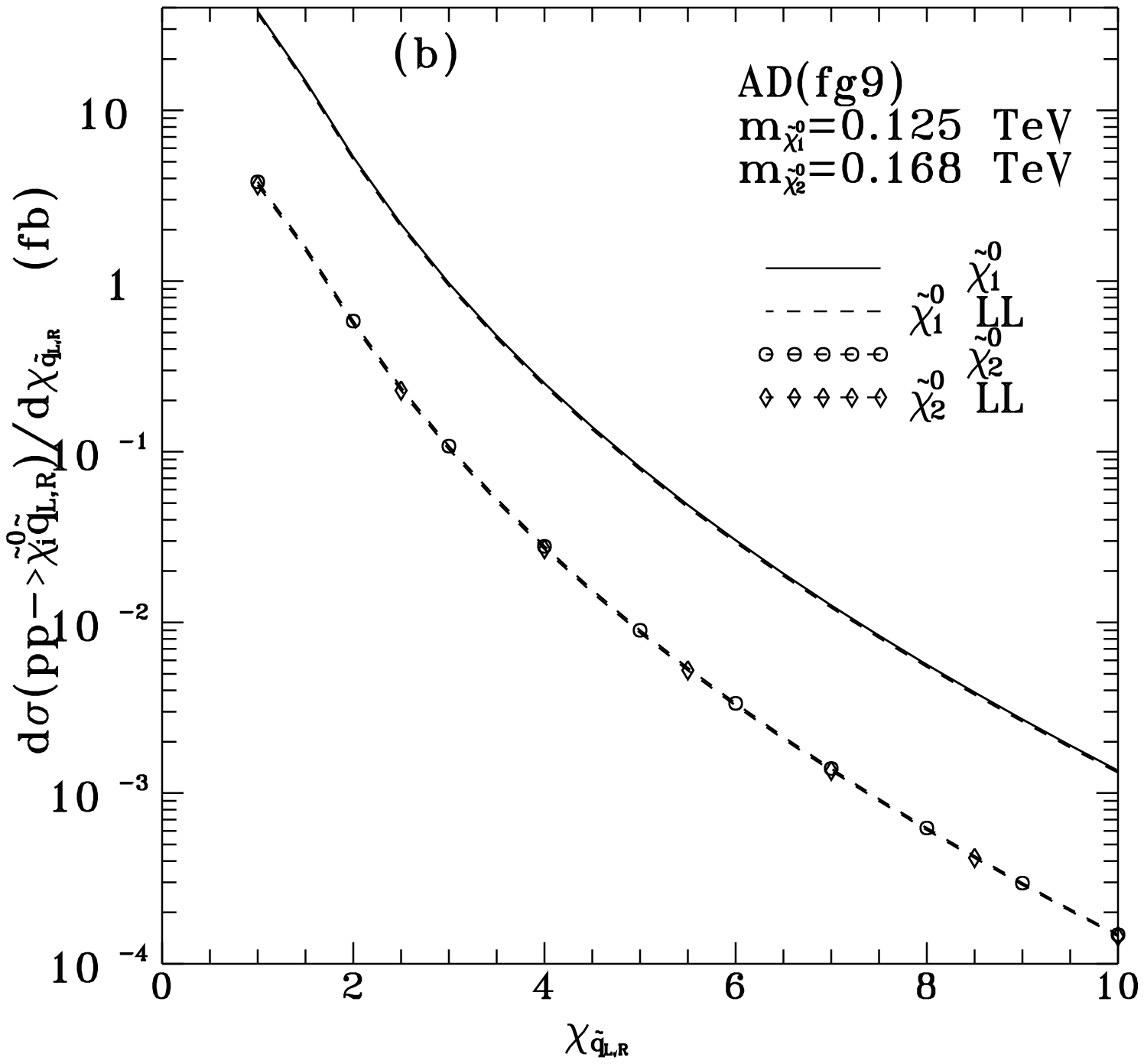,height=7.5cm, width=7.5cm}
\]
\[
\hspace{-0.5cm}\epsfig{file=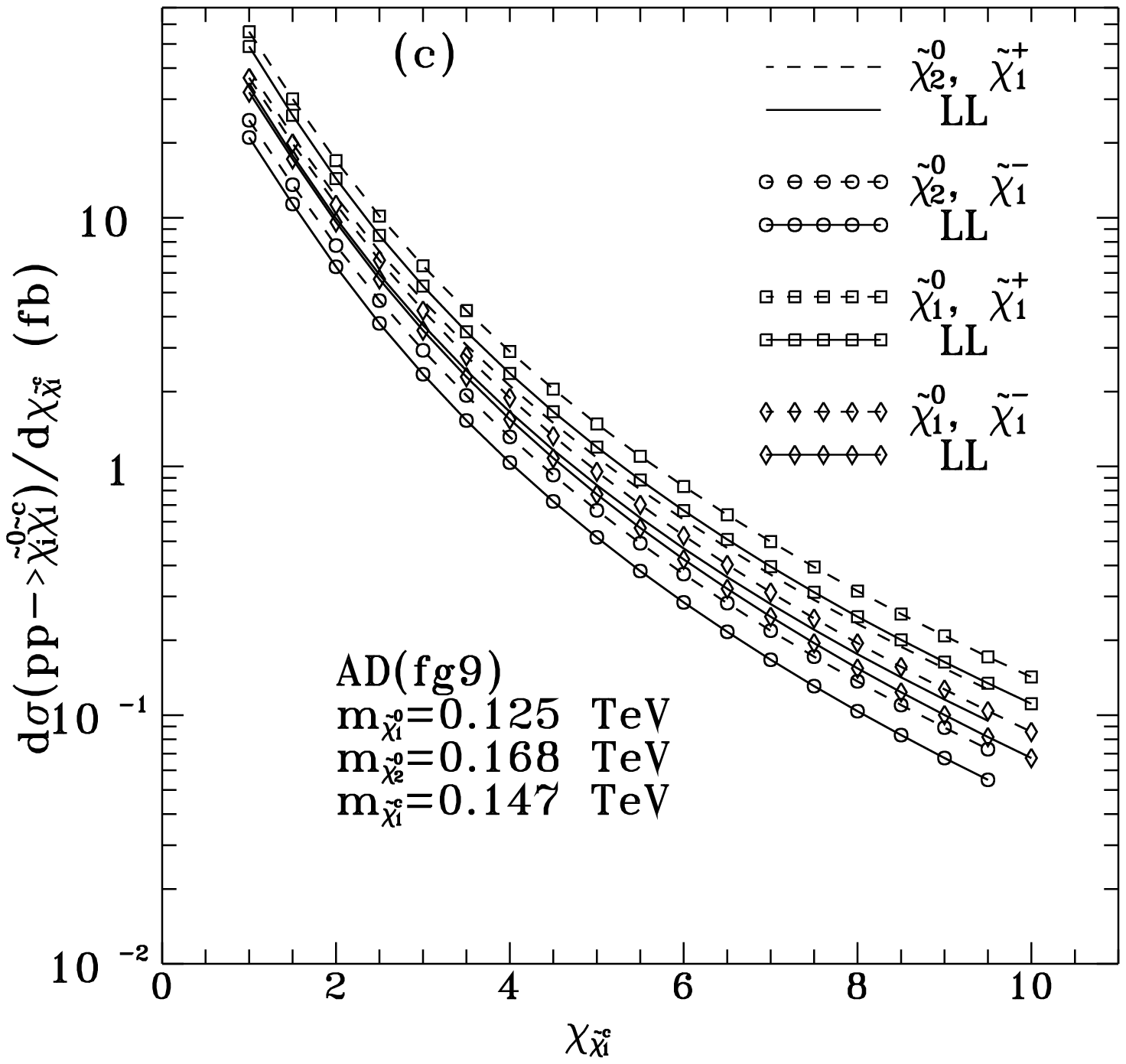,height=7.5cm, width=7.5cm}
\]
\caption[1]{AD(fg9)  $\chi$-distributions for $\tchi^0_{1,2}$ production, in association with
either $\tilde g$  or $\tilde q_{L,R}$ or $\tchi^\pm_1$; ($\tchi^c_j\equiv \tchi^\pm_j$).}
\label{ADfg9-chi-fig}
\end{figure}

\clearpage
\newpage

\begin{figure}[p]
\vspace*{-2cm}
\[
\hspace{-0.5cm}\epsfig{file=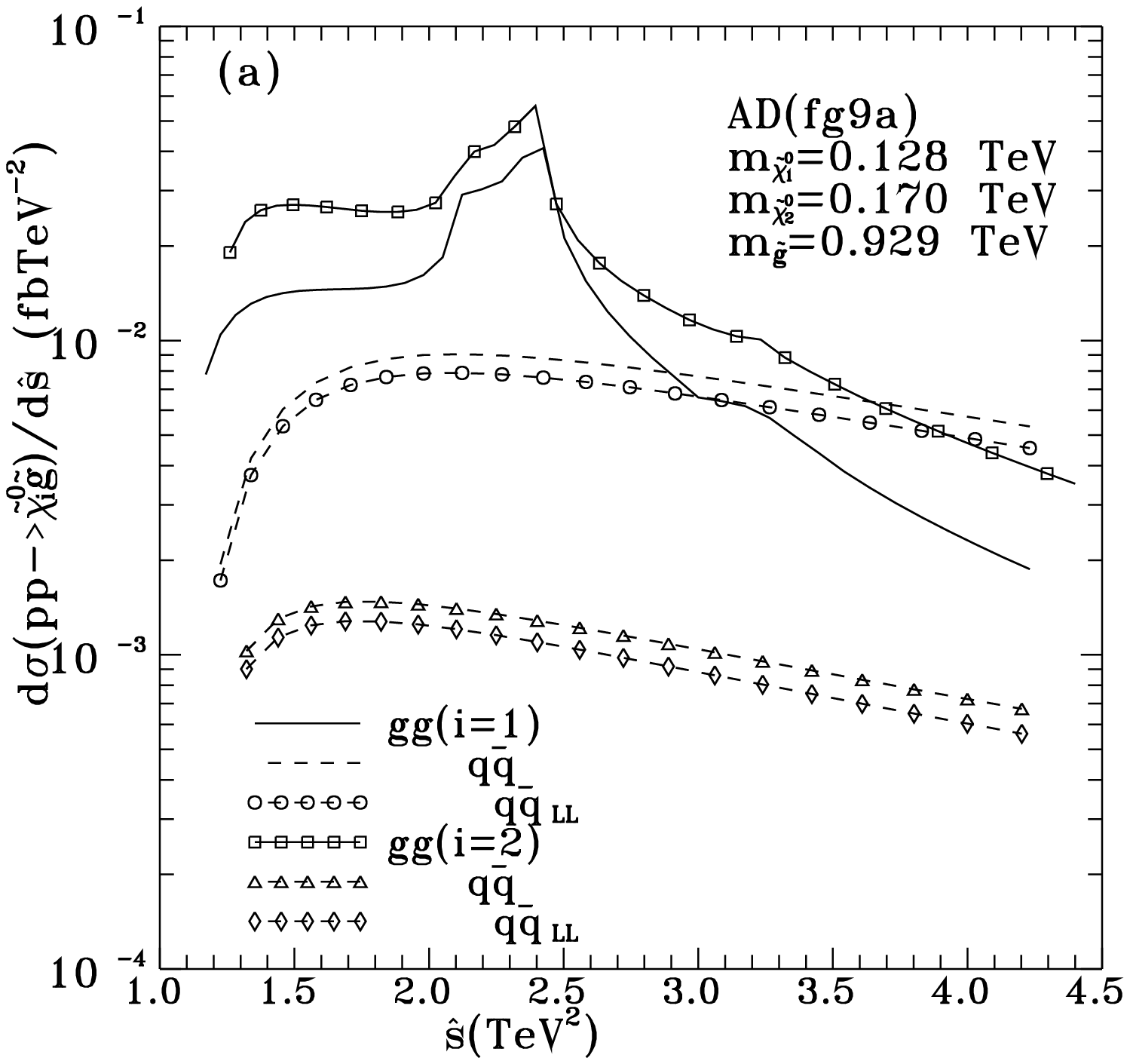,height=7.5cm, width=7.5cm}
\hspace{1.cm}\epsfig{file=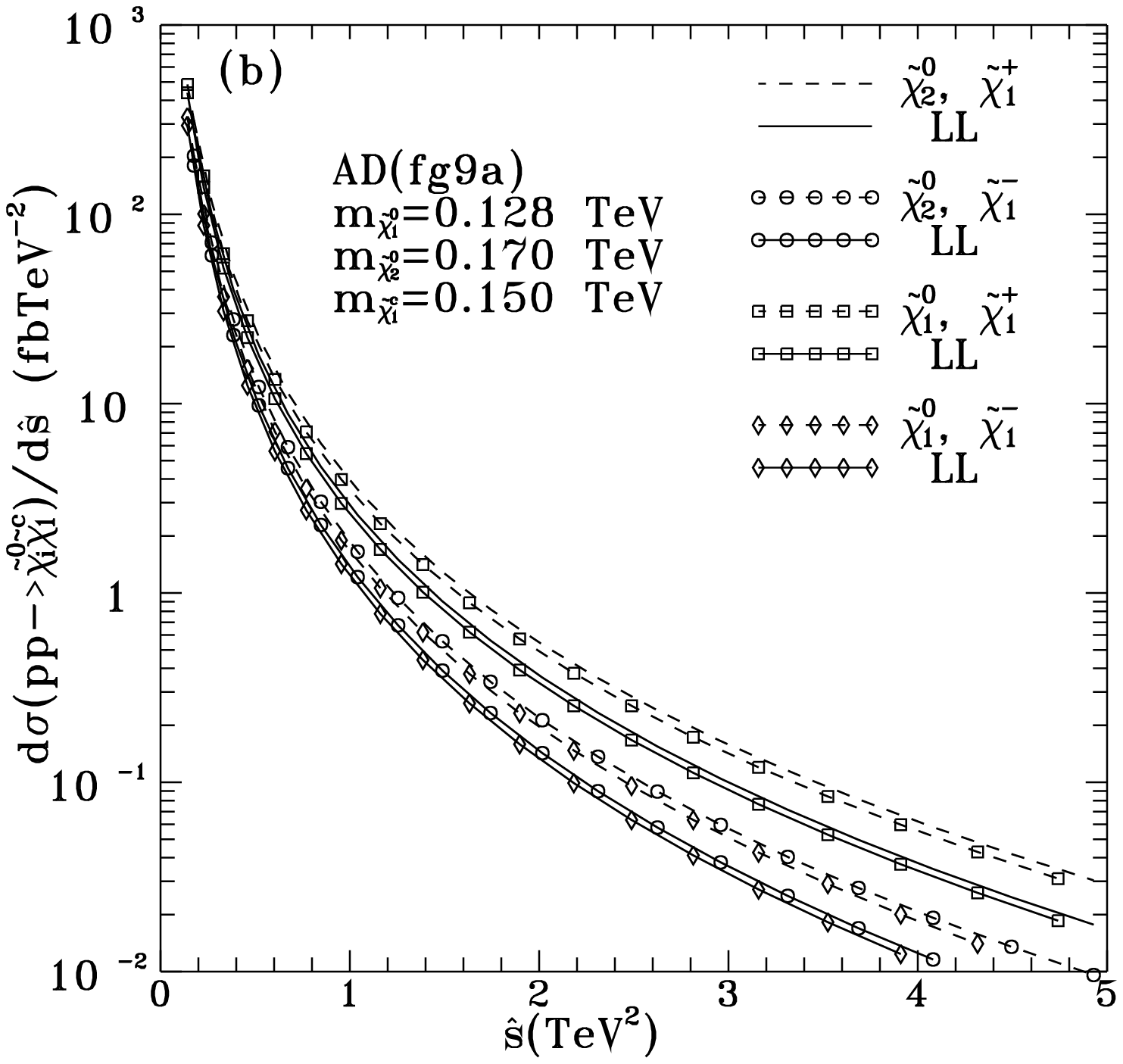,height=7.5cm, width=7.5cm}
\]
\[
\hspace{-0.5cm}\epsfig{file=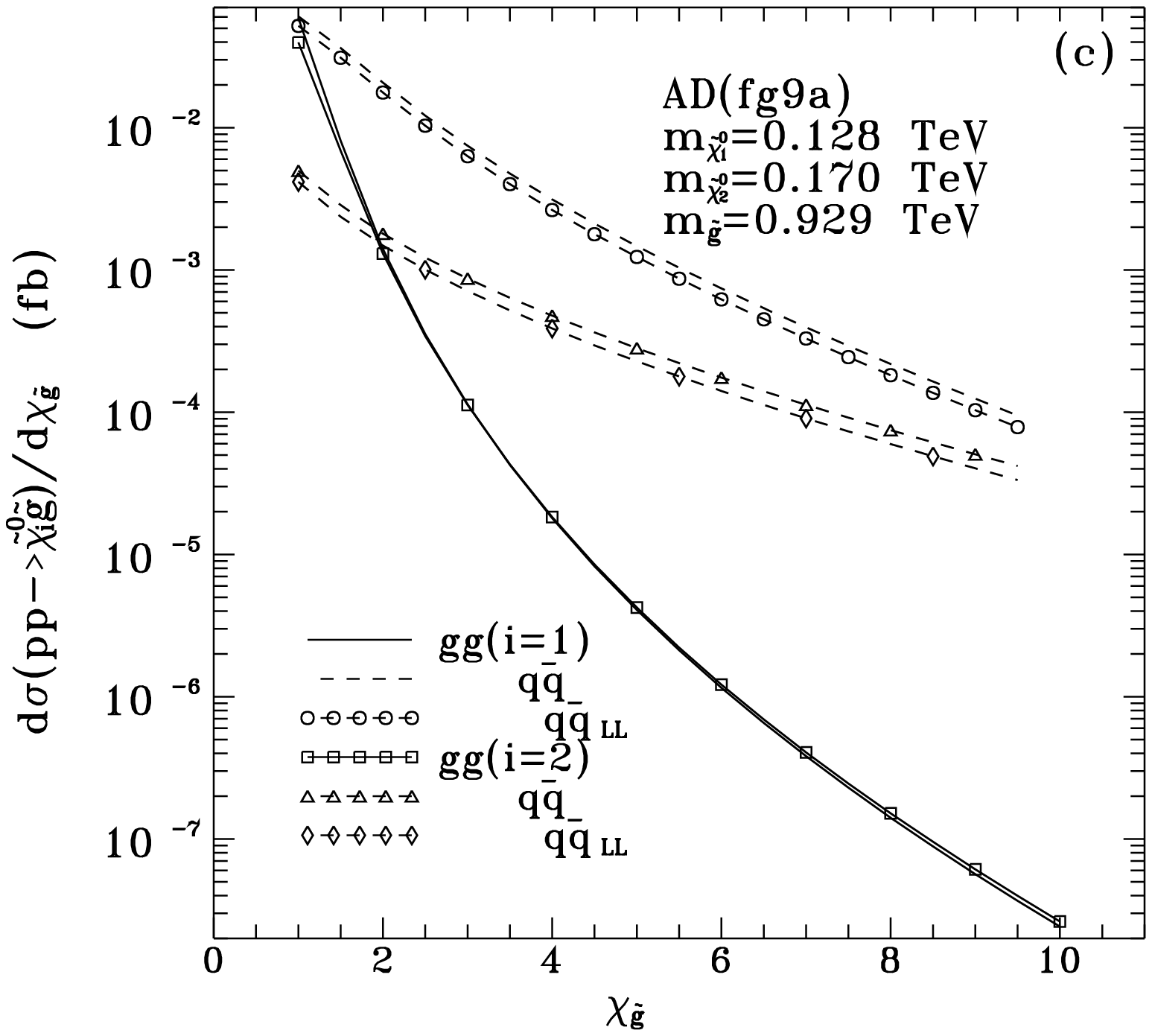,height=7.5cm, width=7.5cm}
\hspace{1.cm}\epsfig{file=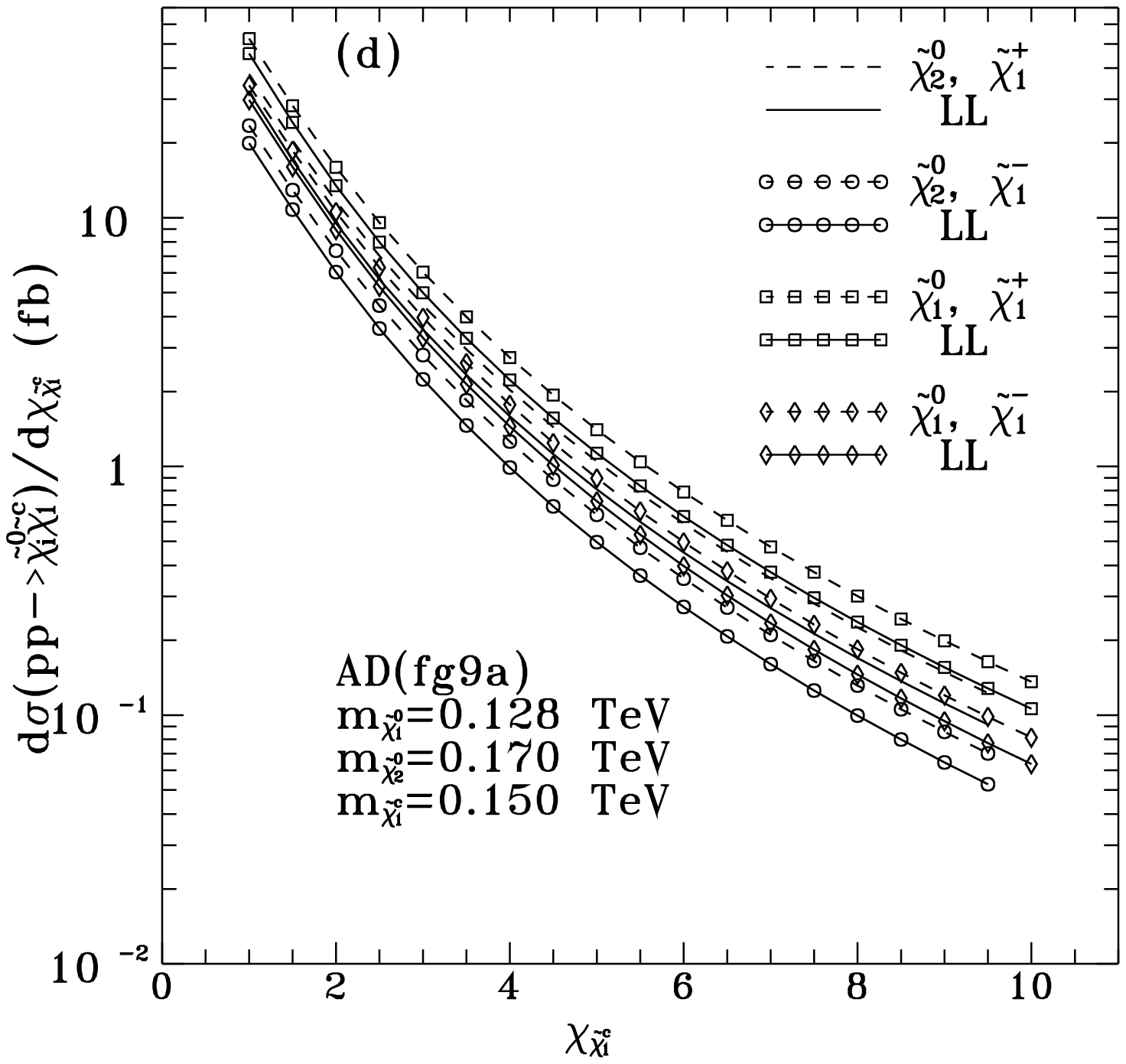,height=7.5cm, width=7.5cm}
\]
\caption[1]{AD(fg9a) $\s$- and $\chi$-distributions for $\tchi^0_{1,2}$ production
in association with either $\tilde g$  or $\tchi^\pm_1$.}
\label{ADfg9a-mass-chi-fig}
\end{figure}

\clearpage
\newpage

\begin{figure}[p]
\vspace*{-2cm}
\[
\hspace{-0.5cm}\epsfig{file=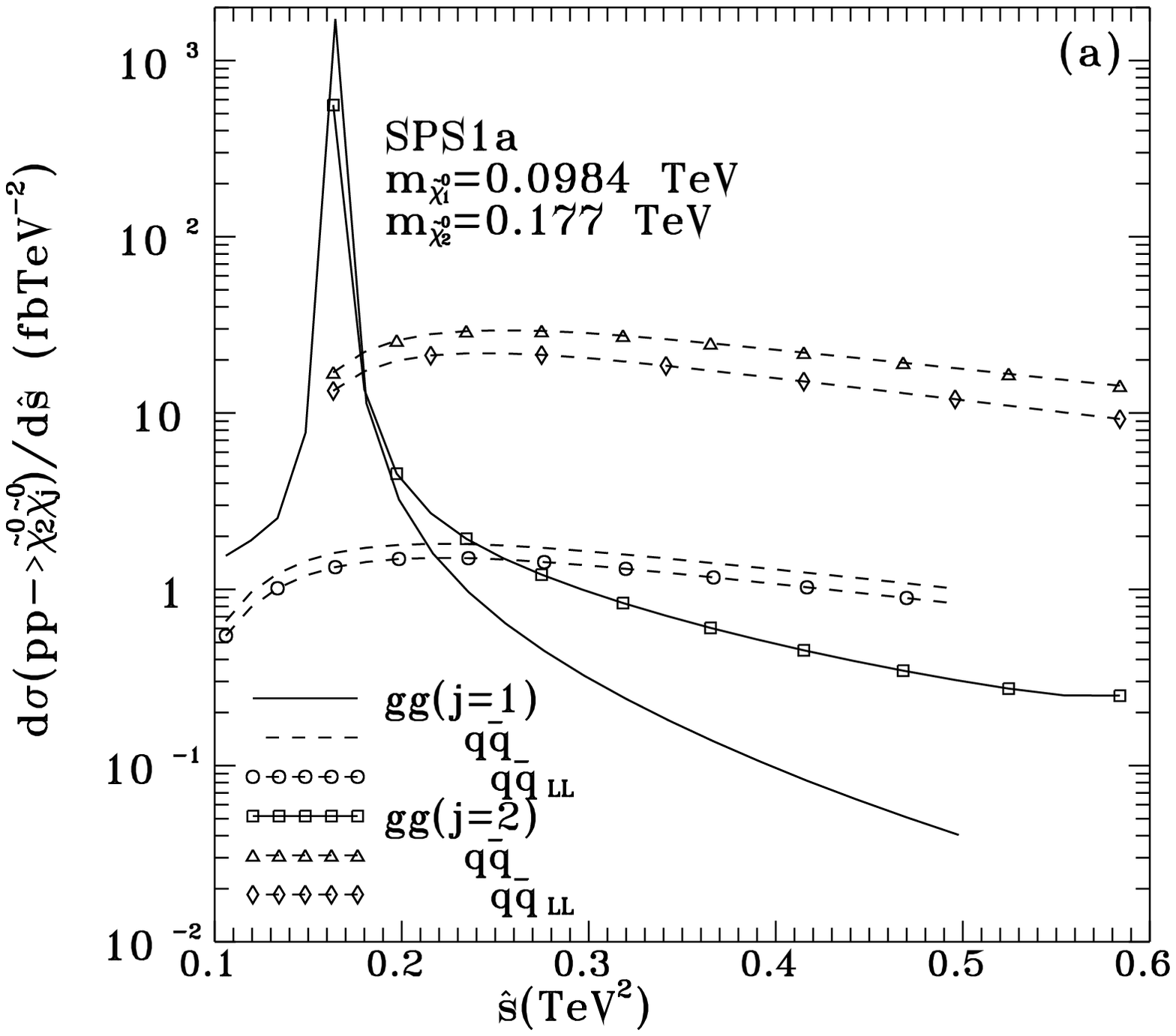,height=7.5cm, width=7.5cm}
\hspace{1.cm}\epsfig{file=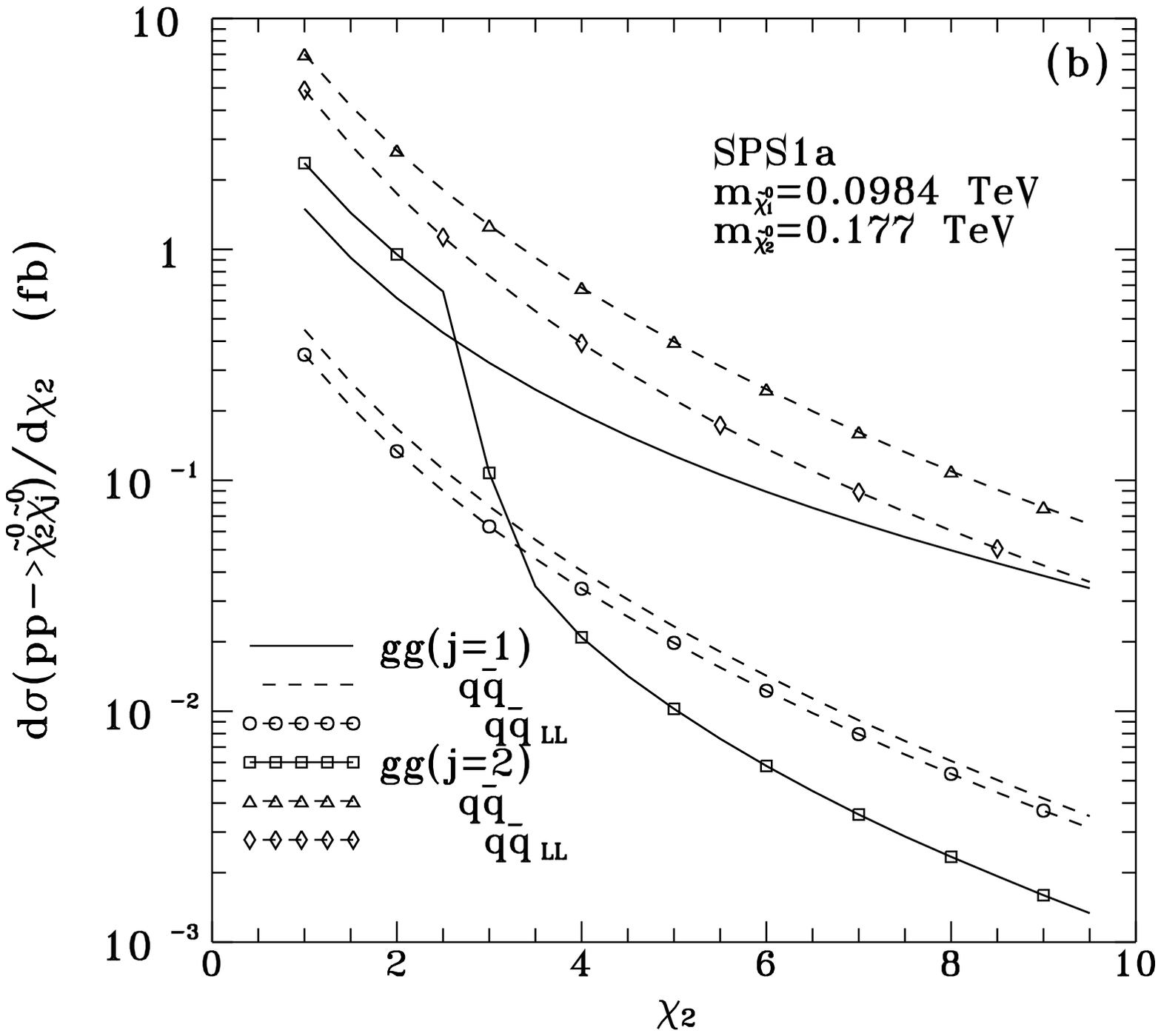,height=7.5cm, width=7.5cm}
\]
\caption[1]{SPS1a  $\s$- and $\chi$-distributions in $\tchi^0_1\tchi^0_j$
production for $j=1$ and $j=2$.}
\label{SPS1a-chi0chi0-fig}
\end{figure}

\begin{figure}[p]
\vspace*{-2cm}
\[
\hspace{-0.5cm}\epsfig{file=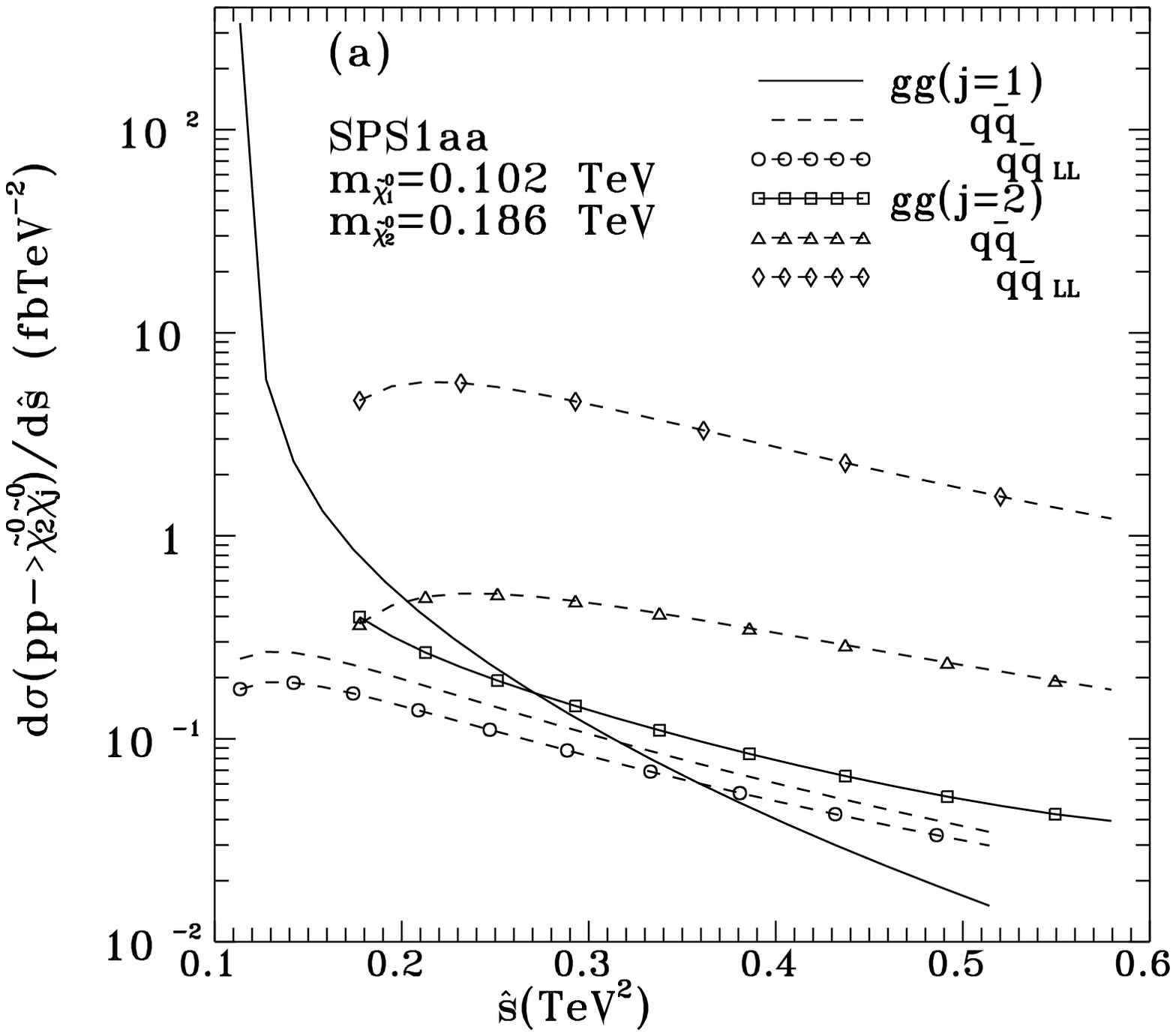,height=7.5cm, width=7.5cm}
\hspace{1.cm}\epsfig{file=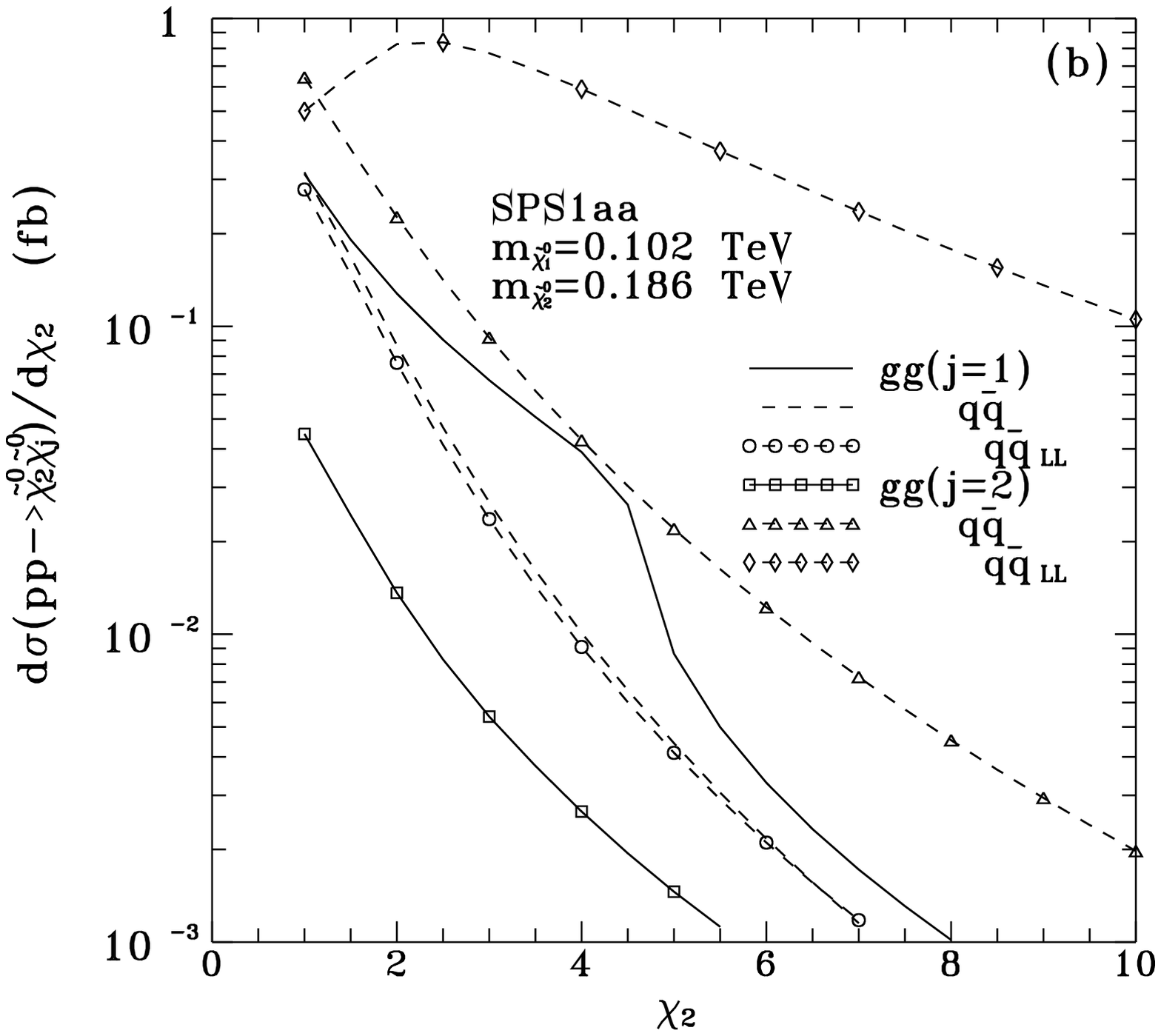,height=7.5cm, width=7.5cm}
\]
\caption[1] {SPS1aa  $\s$- and $\chi$-distributions in  $\tchi^0_1\tchi^0_j$
production for $j=1$ and $j=2$.}
\label{SPS1aa-chi0chi0-fig}
\end{figure}

\end{document}